\documentclass[lettersize,journal]{IEEEtran}
\usepackage{amsmath,amsfonts, amssymb}
\usepackage[hidelinks, colorlinks=true, linkcolor=blue, citecolor=green]{hyperref}

\usepackage[linesnumbered,ruled,vlined]{algorithm2e} 
\usepackage{array}

\usepackage{textcomp}
\usepackage{xcolor} 
\usepackage{titlesec} 
\usepackage{cuted}      
\usepackage{stfloats}
\usepackage{url}
\usepackage{verbatim}
\usepackage{soul}
\usepackage{pifont}
\usepackage{multirow}
\usepackage{graphicx}
\usepackage{cite}
\usepackage{subcaption} 
\usepackage{adjustbox}
\usepackage{booktabs}
\usepackage{threeparttable}
\usepackage{tabularx}
\usepackage{xcolor}
\usepackage{soul}
\usepackage{subcaption} 
\usepackage{caption}
\usepackage{mathtools}
\usepackage{nopageno}
\definecolor{highlightcolor}{RGB}{255, 255, 0}
\sethlcolor{highlightcolor}

\DeclareMathAlphabet{\mathbbold}{U}{bbold}{m}{n}
\RestyleAlgo{ruled}

\usepackage{eso-pic}

\newcommand{\TASEcopyrightnotice}{%
\AddToShipoutPictureFG*{%
  \AtPageLowerLeft{%
    \raisebox{0.28in}{%
      \hspace{0.75in}%
      \begingroup
      \setlength{\fboxsep}{2pt}%
      \setlength{\fboxrule}{0.3pt}%
      \fbox{%
        \begin{minipage}{6.95in}
        \fontsize{6.5}{7.2}\selectfont
 This work has been accepted for publication in IEEE
Transactions on Automation Science and Engineering \textcopyright\ 2026 IEEE. 
IEEE permission is required for uses including reprinting/
republishing, redistribution to servers or lists, resale, creating collective
works, or reuse of copyrighted components. Please cite the published
version when it becomes available.
        \end{minipage}%
      }%
      \endgroup
    }%
  }%
}%
}

\begin{document}

\bstctlcite{IEEEexample:BSTcontrol}

\title{\LARGE \bf FAM-HRI: Foundation-Model Assisted multimodal Human-Robot Interaction Combining Gaze and Speech}
\author{}

\title{\LARGE \bf FAM-HRI: Foundation-Model Assisted multimodal Human-Robot Interaction Combining Gaze and Speech}

\author{Yuzhi Lai$^{1}$,~Shenghai Yuan$^{2}$,~Peizheng Li$^{1,3}$,~Boya Zhang$^{1}$,\\~Benjamin Kiefer$^{1}$,
~Tianchen Deng$^{2,4}$ and Andreas Zell$^{1}$
\thanks{Corresponding Author: \textbf{Andreas Zell}. This work was supported in part by Meta, and we acknowledge their contribution of Meta Glasses for this research.}
\thanks{$^{1}$University of Tuebingen,  Geschwister-Scholl-Platz, 72074 Germany, 
        {\tt\small \{name.surname\}@uni-tuebingen.de}.}%
\thanks{$^{2}$Nanyang Technological University, 50 Nanyang Avenue, Singapore 639798, 
        {\tt\small shyuan@ntu.edu.sg}.}%
        \thanks{$^{3}$
        {\tt\small peizheng.li@mercedes-benz.com}.}%
        \thanks{$^{4}$
        {\tt\small N2308684A@e.ntu.edu.sg}.}
}


\maketitle
 \TASEcopyrightnotice

\pagestyle{empty}
\thispagestyle{empty}
\begin{strip}
\begin{minipage}{\textwidth}\centering
\vspace{-100pt}
\includegraphics[width=0.85\textwidth]{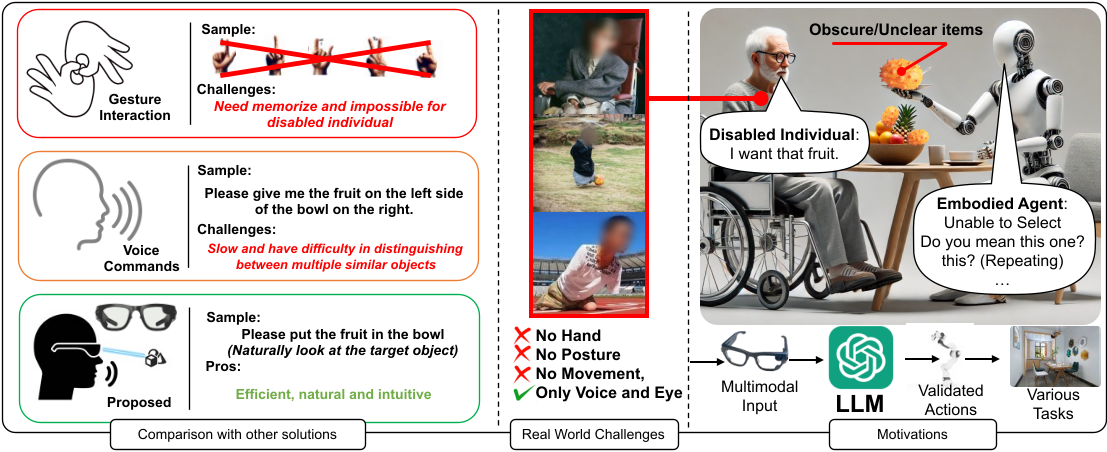}
\vspace{0pt}
\captionof{figure}{Voice-gaze fusion HRI enables efficient, intuitive interaction without memorization, ideal for physically impaired users.  Note: The human and robot images are AI-generated and included solely for illustrative purposes.}
\vspace{-10pt}
\label{figurelabel}
\end{minipage}
\end{strip}

\begin{abstract}

Effective Human-Robot Interaction (HRI) is crucial for enhancing accessibility and usability in real-world robotics applications. However, existing solutions often rely on gesture-only or language-only commands, making interaction inefficient and ambiguous, particularly for users with physical impairments. In this paper, we introduce FAM-HRI, an efficient multimodal framework for HRI that integrates language and gaze inputs via foundation models. By leveraging lightweight Meta ARIA glasses, our system captures real-time multimodal signals and utilizes large language models (LLMs) to fuse user intention with scene context, enabling intuitive and precise robot manipulation. Our method accurately determines the gaze fixation time interval, reducing noise caused by the gaze dynamic nature. Experimental evaluations demonstrate that FAM-HRI achieves a high success rate in task execution while maintaining a low interaction time, providing a practical solution for individuals with limited physical mobility or motor impairments. To support the community, we have      released our system design, algorithms, and solutions at \url{https://github.com/laiyuzhi/FAM-HRI}. 

\end{abstract}

\def\abstractname{Note to Practitioners}
\begin{abstract}
In the field of assistive and service robotics, enabling users with physical disabilities to communicate intentions naturally remains a significant challenge. Existing uni-modal approaches, such as voice-only or gaze-only interaction, often struggle with ambiguity and environmental interference, limiting their reliability in real-world scenarios. Current multimodal systems typically process gaze and speech separately, making it difficult to align visual attention with oral commands accurately. Additionally, many systems rely on bulky hardware or require users to maintain prolonged focus, reducing practicality and comfort.
To address these issues, this paper proposes a foundation-model-assisted multimodal human-robot interaction (FAM-HRI) framework that combines lightweight AR glasses with LLMs to fuse gaze and speech inputs in real time. The system automatically filters noisy gaze signals, aligns them with spoken commands, and generates precise robot actions. Experiments demonstrate that FAM-HRI achieved task success rates exceeding 94\% across multiple scenarios and outperformed all baseline methods in interaction efficiency, with the shortest interaction times in comparative evaluations.
This method is well-suited for applications such as household assistance, and collaborative tasks where hands-free interaction is desirable. In the future, we plan to optimize computing performance and explore strategies to resolve conflicts between gaze and speech inputs, further enhancing usability and adaptability in diverse environments.
\end{abstract}



\section{Introduction}
In recent years, assistive robotics has attracted increasing attention due to its potential to enhance the autonomy and quality of life of individuals with physical disabilities or neuromuscular impairments. Tasks that most people perform effortlessly—like reaching for a cup or opening a drawer—can be daunting for users with limited motor control. Robotic manipulators offer a promising path to automate these actions \cite{c1,c2,c3}. However, realizing this vision hinges on effective and inclusive Human-Robot Interaction (HRI). Traditional interfaces, such as joysticks or touchscreen panels, demand dexterous control and are thus often inaccessible. Voice-controlled systems are more inclusive, yet remain insufficient in accurately interpreting the intentions \cite{lai2026sticky} of the user, particularly when referring to unfamiliar objects or choosing among similar items \cite{c5,c17}. Additionally, popular voice-driven assistants like Amazon Alexa or Tmall Genie lack embodied grounding and often fail in real-world complex tasks. Thus, there is a pressing need for more intuitive, context-aware interaction paradigms that can bridge the gap between human intent and robot execution.

Recent works~\cite{c6,c7,c8,c10} have explored alternative interaction modalities such as eye gaze. Gaze is a natural, low-effort cue that can reveal intent without requiring physical movement.
Early gaze-driven systems have demonstrated feasibility in robot control \cite{c7,c22}, yet they rely on bulky external trackers or head-mounted cameras that impede mobility and resist integration into daily environments \cite{c6,c8,c10}. Moreover, combining gaze with traditional Graphical User Interface (GUI) elements increases interaction time and complexity \cite{c24}. While some approaches attempt to fuse gaze with speech, they typically process the two modalities independently, missing critical cross-modal correlations. As a result, most systems fail to deliver both the precision and ease-of-use needed for everyday assistive HRI, especially in cluttered and dynamic environments of real-world homes.
 Despite its potential, leveraging gaze and speech in HRI introduces several fundamental challenges: (1) Natural gaze behavior is dynamic and noisy: the user’s eyes do not remain fixed but instead shift subtly around the target area \cite{he2024microsaccade}, introducing temporal noise and uncertainty in object selection. Gaze patterns may reflect intention only briefly, complicating efforts to extract meaningful cues. (2) Verbal commands are often ambiguous—especially when users must describe unknown or visually similar items without using precise labels \cite{c17}. (3) Existing systems generally require expensive or immobile hardware setups, which hinder deployment in daily living environments \cite{c6, c8}. (4) Few methods model the temporal alignment between gaze and speech. In real-world scenarios, a user may look at an object moments before or after issuing a verbal command. Without a principled way to synchronize and interpret these signals, errors propagate and degrade the robot’s performance. These challenges underscore the need for lightweight, intelligent, and modality-coordinated interaction frameworks.


To overcome these limitations, we propose \textbf{FAM-HRI}—a \textbf{F}oundation-model-\textbf{A}ssisted \textbf{M}ulti-modal HRI framework that enables intuitive communication for users with limited mobility. 
Our system leverages lightweight wearable glasses from the Meta ARIA Research Kit \cite{c12}, which can simultaneously capture egocentric video, gaze signals, and voice commands. A foundation language model agent processes the spoken command and selects the most semantically aligned gaze period, filtering out irrelevant or noisy frames caused by microsaccades or transient fixations. Another large model agent performs high-level motion planning based on multimodal context and environmental cues. As a result, users can simply glance at their intended object and speak naturally—the system then robustly identifies the target and commands the robot to act. This \textcolor{black}{off-device inference and} joint modeling of gaze and speech enables more precise, adaptive, and accessible HRI, particularly in complex scenes. By reducing ambiguity and minimizing user effort, FAM-HRI bridges the gap between intent and robotic autonomy in assistive settings.

Our main contributions are summarized below:

\begin{itemize}

    \item We propose a multimodal HRI framework designed for wearable, GUI-free interaction. In this framework, speech is used to express complete task intent, while gaze is used for referential disambiguation. Unlike previous approaches \cite{11037823, c7} that require step-by-step user operation or prolonged fixation on GUI, our design enables intuitive and natural interaction without forcing users to continuously control the robot or maintain sustained gaze on target objects or GUI, supporting both high-level (multi-step, multi-object) and low-level (move a fixed distance, rotate) robot actions without teleoperation.
    \item We introduce a reproducible gaze-speech intention alignment and fusion mechanism that integrates multimodal temporal window alignment, adaptive weighting of gaze trajectories, multimodal intention fusion, and deterministic conflict resolution. This mechanism enables robust intention fusion and grounding under noisy and inconsistent gaze-speech inputs, and its individual components are quantitatively validated through ablation studies.
    \item To address head motion and perspective differences between humans and robots, we develop a multi-view alignment module that maps referred targets across changing perspectives. By aligning human and robot observations through feature matching, the proposed method ensures consistent object reference on the tabletop even under large viewpoint variations.
\end{itemize}
\textcolor{black}{Our proposed approach enables robust and natural multimodal interaction for human-robot collaboration using lightweight wearable devices. Compared with the state-of-the-art HRI methods, FAM-HRI achieves higher interaction success rates and faster input processing.}

\section{Related Works}

Recent advances in visual perception  \cite{deng2026voxelized}, SLAM \cite{deng2025mcnslammultiagentcollaborativeneural,li2026doa,deng2025reloc}, and compact robotic hardware \cite{yuan2026peral} have
expanded robot deployment from assistive service scenarios~\cite{c15,c16} to
industrial automation~\cite{c13,c14}. However, natural and reliable intention
communication remains a key bottleneck for human-robot interaction. To facilitate seamless human-robot collaboration, various interaction modalities have gained attention. However, gesture-based systems \cite{c5} and those combining voice and posture \cite{c17, c43} impose physical constraints, making them unsuitable for users with limited mobility or motor impairments. In such cases, language-based interaction emerges as a natural alternative.
However, language-based control presents challenges, including ambiguity in scene descriptions and multiple interpretations of the same instruction, reducing reliability in complex environments \cite{c43}. To address this, some researchers \cite{c21} proposed a vision-language model (VLM)-based dialog system, enabling interactive refinement between the robot and  user \cite{qi2026air}. While this improves robustness, it significantly increases interaction time, limiting real-time efficiency. Alternatively, other studies \cite{c38, c39} explored LLMs for robotic task planning using language input. However, LLMs and VLMs lack pretraining in complex geometric reasoning, leading to failures with multiple similar objects.

\begin{figure*}[t]
      \centering
      \includegraphics[width=0.85\textwidth]{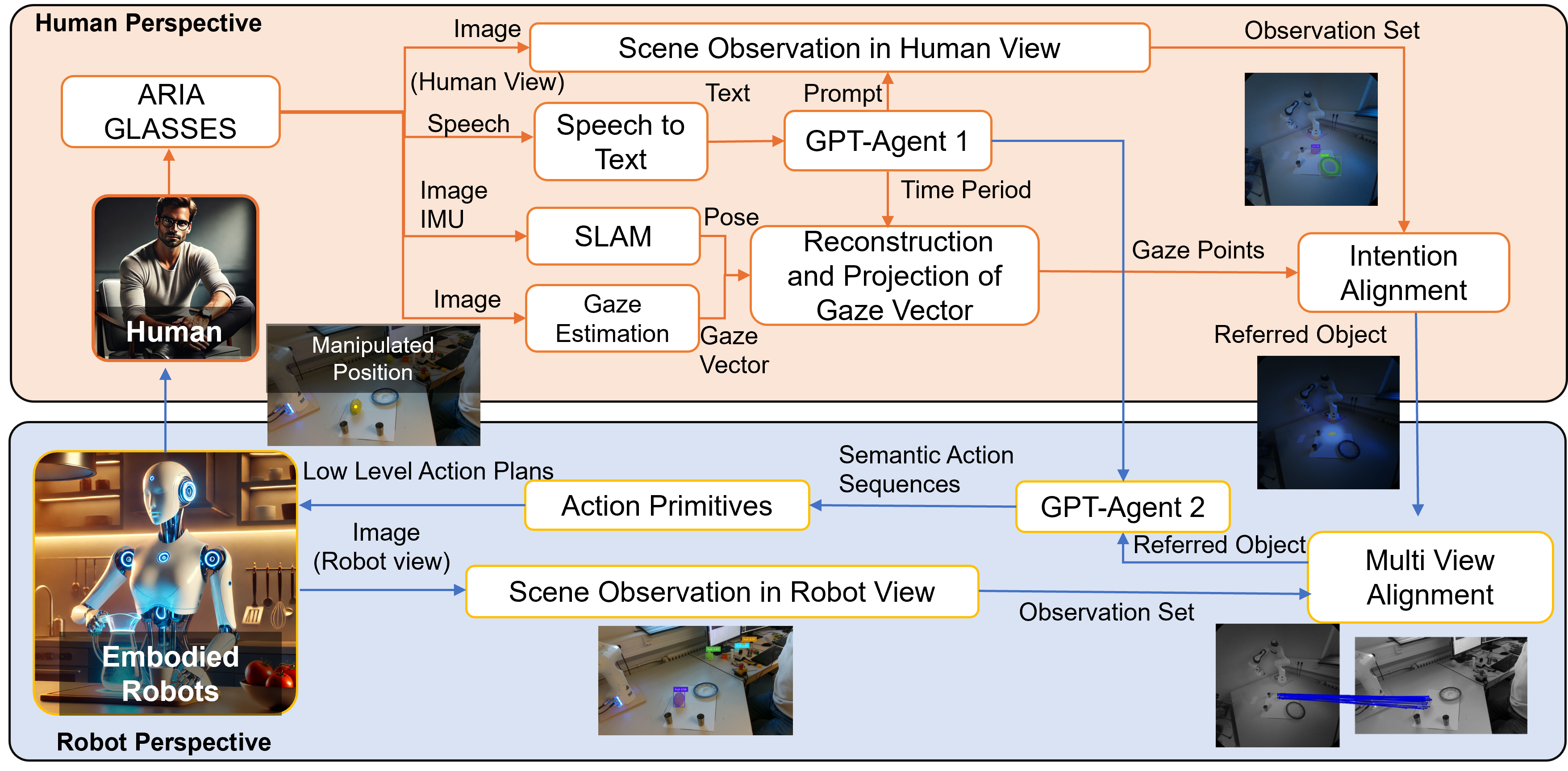}
      \caption{FAM-HRI Framework. Gaze and speech inputs from ARIA glasses are processed by LLM agents to determine the target object and its fixation interval. \textcolor{black}{The framework consists of a human perspective (top) and a robot perspective (bottom). Orange arrows indicate data flow in the human perspective, blue arrows indicate data flow in the robot perspective.} The human view is aligned with the robot view via feature matching, and a planning policy generates parameterized robot actions. \textcolor{black}{(The human and robot images are AI-generated and included solely for illustrative purposes.)}} 
      \label{figure2}
    \vspace{-10pt}
\end{figure*} %

Among alternative modalities, gaze offers a powerful interaction method for individuals with disabilities. Most existing gaze-based assistive devices rely on GUI systems, which enable users to control assistive devices through visual interfaces. Some researchers employ buttons, allowing users to operate electric wheelchairs \cite{c22}, robotic arms \cite{c7, c25}, or mobile robots \cite{c23}.
However, GUI-based systems divert user attention from the external environment, limiting their field of view. An optimal system should enable unrestricted gaze-based interaction by accurately estimating user attention \cite{c24}. FreeView \cite{c24} enables robot teleoperation through a virtual GUI controlled
by the user’s gaze. However, users still need to gaze at the virtual GUI for a long time. Recent work \cite{10806588} focuses on robust gaze-based intention prediction in real-world scenarios. Their method applies OPTICS clustering to filter out redundant gaze points, enabling robust intention prediction. However, this approach presents several limitations. First, it is not integrated with an actual robot control system, and thus lacks validation in closed-loop interaction. Second, it does not consider the alignment between gaze and language input; users can only select target objects through gaze, limiting natural multimodal expression. In \cite{c26}, researchers introduced a gaze-following model that predicts gaze direction and visual attention using RGB images from a robot’s perspective combined with YOLO-based object detection. Similarly, \cite{c27} applies transfer learning to track gaze within predefined regions, allowing the system to determine attention states. This enhances the ability of the robot to interpret user intentions, improving efficiency and naturalness in human-robot collaboration, particularly in industrial settings.
Some recent work \cite{11037823} proposed a human-in-the-loop multimodal fusion method that dynamically adjusts the weights of gaze, gesture, and speech modalities based on user feedback. This adaptive mechanism improves the system's robustness by resolving conflicting inputs during interaction. \textcolor{black}{However, the proposed multimodal fusion does not increase the informational content of the intention inference but merely improves robustness through redundant referencing. Furthermore,} the approach is limited to a fixed set of predefined object categories \cite{cao2024mopa} and supports only simple, hard-coded robot actions. \textcolor{black}{Additionally, it relies heavily on multiple explicit user interventions and repeated user adjustments.} As a result, it lacks the flexibility to engage in natural, open-ended interaction \cite{yang2023av} and cannot support more complex \cite{fan2025structured} or extended task sequences.
However, these approaches also require users to fixate on targets continuously, which feels unnatural compared to typical communication, where the gaze naturally shifts. \textcolor{black}{A multidimensional comparison of representative HRI systems across various modalities is summarized in Tab. \ref{table_comparison related work}.}

\begin{table*}[t]

\caption{\textcolor{black}{Comparison of different related HRI approaches across different modalities.}}
\vspace{-10pt}
\label{table_comparison related work}
\begin{center}
\setlength{\tabcolsep}{0.5pt}
\begin{tabular}{cccccccc}
\hline
\hline
Approach &  Modality & Calibration & Paradigm & Disambiguation& Error Correction & Properties & Task Ability \\ \hline
GesSentence \cite{c5} & Gesture & Hand-eye-sensor & w/o GUI & Pointing & User feedback & Rule based HRI &High/low level task \\
ProgPrompt \cite{c39} & Speech & Hand-eye & w/o GUI & No & No & LLM based reasoning & High level task \\
CaP \cite{c38} & Speech & Hand-eye & w/o GUI & No & No & LLM based reasoning & High level task \\
GIRAF \cite{c43} & Gesture+speech & Hand-eye & w/o GUI & pointing & No & Fusion by LLM & High level task \\
NVP-HRI \cite{c17} & Gesture+speech & Hand-eye & w/o GUI & pointing & LLM reasoning & Fusion by LLM & High level task \\
InteractiveDialog \cite{c21} & Speech & Hand-eye & w/o GUI & Dialog & No & User guided by VLM & Simple pick-place \\
MR-HRI \cite{c7} & Gaze+speech & Hand-eye & Physical GUI & GUI & No & Independent modality & Simple pick-place \\
FreeView \cite{c24} & Gaze & Irrelevant & Virtual GUI & Irrelevant & Irrelevant & Teleoperation & Low level task \\
OpticGaze \cite{10806588} & Gaze+speech & Sensor-environment & w/o GUI & Gaze & Irrelevant & Target selection only & Irrelevant  \\
GazeEEG \cite{c25} & Gaze+EEG & Hand-eye & Physical GUI & No & No & Independent modality & Simple pick-place \\
HIL-HRI \cite{11037823} & Gaze+speech+gesture & Hand-eye-sensor & w/o GUI & User intervention & User intervention & Modal redundancy & Simple pick-place \\
GazeWheelChair \cite{c22} & Gaze & Sensor-hand & Physical GUI & Irrelevant & Irrelevant & Teleoperation & Low level task \\
Ours & Gaze+speech & Hand-eye & w/o GUI & Gaze & User intervention & Modal complementarity & High/Low level task \\
\hline
\hline
\vspace{-10pt}
\end{tabular} 
\end{center}
\footnotesize{ 

}
\vspace{-10pt}
\end{table*}

Existing HRI approaches \cite{c17,yuan2025starc,lai2025seer} remain inadequate for general use, particularly for individuals with difficulties in language commands or physical movement. Integrating LLMs with strong reasoning capabilities \cite{deng2025best3dscenerepresentation} offers potential for enhancing multimodal data processing across various HRI applications, including social robots \cite{c17, c43} and robot navigation \cite{c29}.
Recently, researchers introduced LaMI \cite{c31,xu2026deployable} to explore how LLMs can improve multimodal HRI by reasoning over diverse inputs. While LaMI provides a strong backend reasoning capability, it primarily focuses on symbolic-level interpretation based on human pose estimation, dialogue history and scene detection. This approach overlooks frontend measurement uncertainties \cite{deng2025mne} and other external factors \cite{yuan2026unsupervised} affecting perception \cite{yuan2026adaptive} and decision-making \cite{li2026aeos} in real-world interactions \cite{liu2026global,yuan2026peral}. For example, LaMI infers the user's attention direction based on head orientation to determine whether the user is referring to the robot or the workspace, rather than determining the specific object to which the user is visually attending.


Unlike previous HRI approaches \cite{c17} that struggle with frontend uncertainties and real-world perception challenges, our method leverages lightweight ARIA glasses to capture high-precision gaze data and speech signals. Using LLM-based reasoning, we accurately determine the most probable gaze fixation period, minimizing noise from gaze dynamics. Additionally, a second LLM agent analyzes scene context and user intent to generate optimal robot actions, ensuring more efficient and natural HRI.


\section{Problem Definition}
Our proposed system integrates environmental perception from both human and robot views to facilitate effective collaboration. The \textbf{human view} is mainly used for intention fusion, capturing user-centric cues, while the \textbf{robot view} focuses on manipulation, allowing precise interaction with objects. To structure multimodal signals for foundation model reasoning, we formalize sets of geometric frames and control parameters. For comprehensive definitions of the mathematical notations used throughout our paper, please refer to Appendix \ref{symbols}.

\subsection{Definitions of Frames and Transformations} 
To ensure consistency, we define the reference frames used in this paper. The \textbf{robot base frame} $\prescript{r}{}{(\cdot)}$ serves as the primary reference for the robot, while the \textbf{robot camera frame} $\prescript{c}{}{(\cdot)}$ is used for manipulation. 
The \textbf{ARIA glasses camera frame} $\prescript{gc}{}{(\cdot)}$ provides a first-person user-centric perspective. \textbf{The ARIA pupil frame} $\prescript{gp}{}{(\cdot)}$, centered at the midpoint of the glasses' bridge, serves as the reference for gaze vector estimation. The transformation $T \in \mathbb{SE}(3)$ denotes a rigid body transformation in the special Euclidean group. Specifically, the transformation matrix ${}^{gc}T_{gp}$ maps points from the ARIA pupil frame to the ARIA camera frame, with parameters provided by the manufacturer\footnote{\scriptsize \url{facebookresearch.github.io/projectaria_tools/}}.


\subsection{Formulation of Human View Representation}
To formally characterize the human view inputs of ARIA glasses, we define the human input state vector as
$\mathcal{Z}_{H} = (\mathcal{S}, \mathcal{U}, \mathcal{G})$.  
Here, $\mathcal{S} \in \Sigma^*$ denotes the transcribed speech sequence. The audio is recorded at a 48 kHz sampling rate using the built-in microphone of the ARIA glasses and is converted into text via Fast Whisper \cite{c32}, with each word timestamped. Six embedded microphones improve speech clarity by suppressing background noise and improving transcription accuracy even in noisy environments. 
$\mathcal{U} \in \mathbb{R}^{H_1 \times W_1 \times (t_n - t_1)}$ represents the RGB image sequence captured over the interval $(t_n - t_1)$ by the user-centric camera (Resolution: $1408 \times 1408$, 10 FPS).  
$\mathcal{G} \in \mathbb{R}^{H_2 \times W_2 \times (t_n - t_1)}$ corresponds to the grayscale eye-tracking image sequence (Resolution: $320 \times 240$ per eye, 10 FPS), independently capturing the left and right eyes for gaze estimation.

The human-view inputs are utilized to infer the referred object, represented as  
$\overline{\mathcal{Z}_g} = (\overline{{p}_{g}}, \  \overline{\beta_{g}}, \  \overline{M_{g}}), $ 
which encodes its spatial and visual properties. Here, $\overline{{p}_{g}} \in \mathbb{R}^2$ denotes the 2D position of the referred object, $\overline{\beta_{g}} \in \mathbb{R}^4$ defines its bounding box, and $\overline{M_{g}} \in \mathbb{R}^{H_1 \times W_1}$ represents its segmentation mask. $\overline{{p}_{g}}$ is defined as the midpoint of the $\overline{\beta_{g}}$.

\subsection{Formulation of Robot View Representation }

The robot camera captures the scene from the robot perspective, with input $\mathcal{C} \in \mathbb{R}^{H_3 \times W_3}$ (Resolution: 1280 $\times$ 720) representing the image of the RGBD camera. The goal of the robot is to correlate the same referenced object within the robot view and the human view. This reference object in robot view is denoted as $\overline{\mathcal{Z}_{r}}=( \overline{{p}_{r}}, \  \overline{\beta_{r}}, \  \overline{M_{r}} )$. Similar to the human view, $\overline{{p}_{r}}$, $\overline{\beta_{r}}$ and $\overline{M_{r}}$ encode the object's spatial and visual properties within the robot's perspective.

\subsection{Formulation of Control System}

The control system is responsible for generating and executing actions based on the intentions of the human and robot views. The system is manipulated with a predefined set of parameterized action primitives  $\mathcal{A}(\Theta) = \left \{ a_i(\theta_i) \mid i \in \mathbb{R} \right \}$ and follows a parameterized planning policy to ensure adaptive and context-aware manipulation. 
In $\mathcal{A}(\Theta)$, each primitive action $a$ is parameterized by 2D position, object category, translation distance, and rotation angle. These action primitives form low-level and high-level control commands, enabling the robot to execute pick-up, placing, pushing, moving, and rotating actions in a structured manner.

The execution strategy follows a parameterized planning policy $\pi$, which is denoted as:
\begin{align}
   \pi : X \mapsto \ \mathcal{A}\times \Theta.
    \label{eqpolicy}
\end{align}

As the input to the policy $\pi$, $X$ represents the state space, incorporating scene observations and task-specific constraints. $\mathcal{A} \times \Theta$ denotes action-parameter pairs. $\Theta$ represents the parameters of action primitives. The policy generation will be discussed in Sec. \ref{policy}.

\section{Methodology}

Our system enables multimodal HRI by integrating gaze, language, and visual perception across both human and robot viewpoints to achieve accurate intention inference and action execution. The methodology is structured into three key components: Human View Intention Fusion, Multi-View Intention Alignment, and Planning Policy Generation. Our proposed system diagram is shown in Fig. \ref{figure2}. 


\subsection{Human View Intention Fusion}
When a command is given, the human’s gaze naturally shifts
and fixates on the target only at specific moments within the
command, making direct fixation-based selection unreliable. Our system aims to minimize this noise and achieve robust intention fusion by aligning gaze and language.
\subsubsection{LLM-Agent for Human View Command Processing}
To effectively fuse gaze and language for intention inference, we leverage a LLM-based agent (GPT-4o, frozen weight 2024-08-06) to analyze the spoken command in text and determine the object or location referenced and the most relevant fixation period. This LLM-Agent takes $\mathcal{S}$ as input and produces the following structured output as:
\begin{align}
    \mathbf{O}_1 = \left \{  \mathcal{L},\ c_{target},\ \Delta t   \right \},
    \label{o1}
\end{align}
where $\mathcal{L}$ defines the target properties, with possible values ``\textit{object}" or ``\textit{position}". The variable $c_{target}$ specifies the target object category in the command, later utilized for scene observation in both human and robot views.  If the referenced target is a general category, the output is ``\textit{stuff}"; if it specifies a position without an object, it is labeled ``\textit{position}"; otherwise, it corresponds to a specific object category.  
$\Delta t$ indicates the time period of the word spoken by the user, determined by Fast Whisper \cite{c32}, during which the user is most likely fixating on the target object or location. For instance, given the command: ``please put the apple there on the table". The system interprets it as: $\mathbf{O}_1 =  \left \{ (object, \ position),\ (apple,\ table),\ (\Delta t_{apple}, \ \Delta t_{there} ) \right \} $. This structured output aligns gaze fixation with spoken language to enhance intention inference. \textcolor{black}{The prompt is shown in Appendix. \ref{PromptA}.}

\subsubsection{Scene Observation in Human View}

We define a scene observation set in human view $\mathcal{Z}_g$ to describe target object categories spatial and visual properties as $\mathcal{Z}_g = \left \{ c_{target}, \ \{{p}_{g_i}, \ \beta_{g_i}, \ M_{g_i} \}_{i=1}^{N} \right \}$.
In this set ${p}_{g_i}\in\mathbb{R}^2$, $\beta_{g_i} \in \mathbb{R}^4$ and $M_{g_i} \in \mathbb{R}^{H_1\times W_1}$ are the 2D position, bbox and mask of the target object $i$ and $c_{target}$ is the category for target object from the first LLM-Agent. ${p}_{g_i}$ is defined as the midpoint of the $\beta_{g_i}$. A total of $N$ objects of this category are observed. In our proposed approach, Grounding DINO~\cite{c35} and SAM 2~\cite{c36} perform object detection, classification, and segmentation with prompt $c_{target}$ from Eq. \ref{o1}.
\subsubsection{3D Gaze Estimation and Reconstruction}
For 3D gaze estimation, the ARIA glasses incorporate two eye-tracking cameras. We employ the SocialEye model\footnote{\scriptsize \url{github.com/facebookresearch/projectARIA_eyetracking}} to predict the 3D gaze vector ${}^{gp_{t_i}}\mathbf{P}_{{t_i}} \in \mathbb{R}^3$ at time $t_i$ in the pupil frame of the glasses. To reconstruct the 3D gaze vector within $\Delta t$ at time $t_{i+n}$, denoted as ${}^{gc_{t_{i+n}}} \mathbf{P}_{{t_{i+n}}}$, onto the frame $\prescript{gc}{}{(\cdot)}$ at time $t_{i}$, denoted as ${}^{gc_{t_{i}}} \mathbf{P}_{{t_{i+n}}} \in \mathbb{R}^3$, and then project them onto image plane ${}^{{t_i}}\mathcal{U}$ at $t_i$ as gaze point ${}^{gc_{t_{i}}} {p}^{gaze}_{{t_{i+n}}} \in \mathbb{R}^2$  we use the transformations:
\begin{align}
    {}^{gc_{t_{i+n}}} \mathbf{P}_{{t_{i+n}}} &= \textcolor{black}{{}^{gc}T_{gp_{t_{i+n}}}} {}^{gp_{t_{i+n}}}\mathbf{P}_{{t_{i+n}}}\notag \\
    {}^{gc_{t_{i}}} \mathbf{P}_{{t_{i+n}}}  &=({}^{s}T^{}_{gc_{t_{i}}})^{- 1}({}^{s}T^{}_{gc_{t_{i+n}}}) {}^{gc_{t_{i+n}}} \mathbf{P}_{{t_{i+n}}}  \\
    {}^{gc_{t_{i}}} {p}^{gaze}_{{t_{i+n}}}&=\mathcal{K} {}^{gc_{t_{i}}} \mathbf{P}_{{t_{i+n}}} \notag
\end{align}

The user-centric camera intrinsic matrix  
$\mathcal{K} \in \mathbb{R}^{3\times3}$  
is obtained from the  ARIA glasses factory calibration data.  
The transformations ${}^{s}T^{}_{gc_{t_{i}}}$ and  
${}^{s}T^{}_{gc_{t_{i+n}}}$ represent the pose of the glasses'  
camera at timestamps $t_i$ and $t_{i+n}$, respectively. \textcolor{black}{${}^{gc}T_{gp_{t_{i+n}}}$ denotes the transformation matrix between pupil  and  camera frame at time $t_{i+n}$.}  
These transformations are estimated using ORB-SLAM3 \cite{c33},  
which integrates an IMU (800Hz) and a Monochrome Scene Camera  
(Resolution: 640$\times$480, 20 FPS) for monocular-inertial SLAM. The index $n$ is defined as  
$n \in (0, \ 1, ..., \left \lfloor \Delta t / 20 -1  \right \rfloor),$
where 20 corresponds to the monochrome camera's frame rate.  
To reduce noise and computational redundancy,  
our system extracts a single gaze point per head pose,  
avoiding scene observation generation in human view  
for each frame within $\Delta t$.

\subsubsection{Human View Intention Alignment}
\label{Human View Intention Alignment}
After projecting gaze vectors onto a single image as ${p}^{gaze}_{{t_{i+n}}}$ and generating the corresponding scene observation $\mathcal{Z}_g$, the next goal is to fuse gaze and language to infer the specific referred object $\overline{\mathcal{Z}_g}$ of the user. \textcolor{black}{In our proposed approach, this intention fusion is determined by a gaze trajectory filter algorithm $f(\cdot)$ with respect to ${}^{gc_{t_{i}}} {p}^{gaze}_{{t_{i+n}}} $ and ${p}_{g_i}$ as: }

\begin{align}
    \overline{\mathcal{Z}_{g}} = f({}^{gc_{t_{i}}} {p}^{gaze}_{{t_{i+n}}},\ {p}_{g_i})
\label{zg}
\end{align}


\subsubsection{\textcolor{black}{Gaze trajectory filter function}}
\label{Gaze trajectory filter function}
  \begin{figure*}[t]
      \centering
      \includegraphics[width=0.85\textwidth]{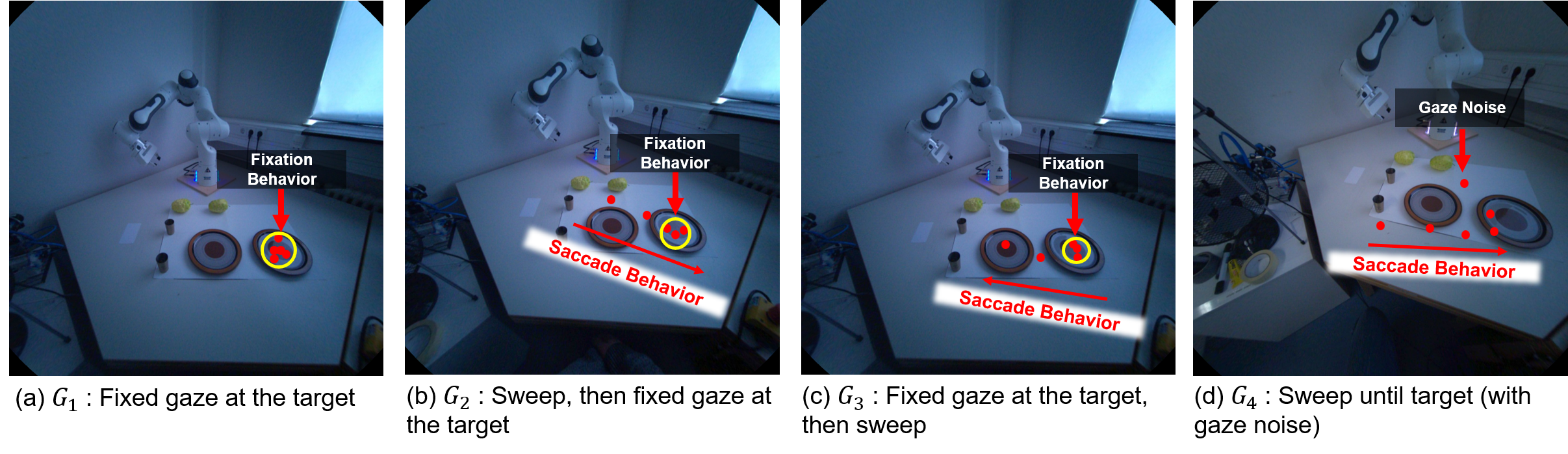}
      \vspace{-10pt}
      \caption{\textcolor{black}{Typical gaze trajectory patterns when selecting intent objects. Gaze trajectories are composed of fixation and saccade behaviors. $G_1$: Fixed gaze at the target; $G_2$: Sweep, then fixed gaze at the target; $G_3$: Fixed gaze at the target, then sweep; $G_4$: Sweep until near the target, due to inertia, the later gaze point may drift away from the target object (gaze noise).}}
      \label{G1-3}
      \vspace{-10pt}
  \end{figure*}

\textcolor{black}{Prior studies \cite{yarbus2013eye, mohamed2024review} have shown that when users have a latent intention to select a target object, their gaze behavior exhibits systematic changes in gaze trajectories. In this situation, gaze trajectories are composed of fixation and saccade behaviors, exhibiting diverse characteristics when approaching, focusing on, or departing from target objects, reflecting four patterns as shown in Fig. \ref{G1-3}. }

\textcolor{black}{Based on the four typical gaze trajectory patterns ($G_1$–$G_  4$), we design a filtering algorithm that is robust to both fixation and saccade behaviors. We first apply OPTICS clustering \cite{ankerst1999optics} on the 2D gaze samples with parameters 
$\text{MinPts} = 3$ and $\epsilon = \infty$. This avoids predefining a fixed neighborhood radius and does not require specifying the number of clusters. We then enforce temporal continuity by removing points that are not contiguous in time and only keeping segments containing at least three samples. For each remaining segment, we compute a minimum enclosing circle (MEC) to represent the spatial extent of visual attention. If at least one valid MEC exists ($G_{1-3}$), the referred object $\overline{\mathcal{Z}_{g}}$ is selected as the one which is closest to the MEC center $o$ i.e. $\textcolor{black}{\overline{\mathcal{Z}_{g}} = \mathop{\arg\min}\limits_{{p}_{g_i}\in \mathcal{Z}_g} \left \| o -{p}_{g_i} \right \| }$.}

\textcolor{black}{If no valid MEC can be formed ($G_4$), we estimate the referred object using a recency-weighted distance accumulation: }
\begin{align}
\textcolor{black}{\overline{\mathcal{Z}_{g}} = \mathop{\arg\min}\limits_{{p}_{g_i}\in \mathcal{Z}_g} \sum_{n=0}^{N-1} e^{\alpha( n-N)} \left \| {}^{gc_{t_{i}}} {p}^{gaze}_{{t_{i+n}}} -{p}_{g_i} \right \| }    
\end{align}

\begin{figure}[t]
      \centering
      \includegraphics[width=0.40\textwidth]{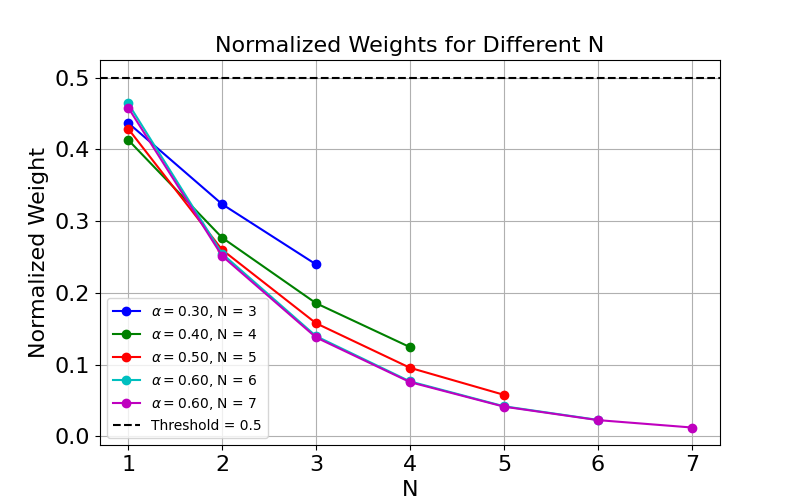}
     \vspace{-10pt}
      \caption{Normalized Weights under Different $N$.}
      \vspace{-10pt}
      \label{weight}
\end{figure}

\textcolor{black}{Where $N$ denotes the number of gaze points in $\Delta t$. We define $\alpha$ as: $\alpha = 0, \text{if } N=2,$ otherwise $\alpha = \min(0.60, \ 0.1N)$. The rationale is shown in Fig. \ref{weight}. When $N = 2$, we disable the exponential factor to avoid over-emphasizing the last sample in a very short window. For longer windows and to avoid the gaze noise (shown in Fig. \ref{G1-3} (d)), $\alpha$ increases linearly with $N$ to strengthen the preference for later gaze samples, while being capped at 0.60 to prevent the sample from dominating the score. $\alpha$ is selected via sensitivity analysis as shown in Appendix \ref{Sensitivity}. }

\subsection{Multi-View Intention Alignment}

Our system incorporates two cameras: one user-centric camera on the glasses and a robot camera. Through human view intention fusion, the referred object $\overline{\mathcal{Z}_{{g}}}$ is determined, while the robot also needs to determine the referred object $\overline{\mathcal{Z}_{{r}}}$. Instead of relying on complex extrinsic calibration between the robot camera and the user-centric camera on the glasses, we implemented a feature-based alignment approach. We first obtain the scene observation of the robot view with the same process used for the human view, generating the scene observation set $\mathcal{Z}_r =\left \{ c_{target}, \ \{{p}_{r_i}, \ \beta_{r_i}, \ M_{r_i} \}_{i=1}^{N} \right \}$.
To establish a correspondence between the human and robot viewpoints, we restrict feature matching from keypoints inside the bounding box of $\overline{\mathcal{Z}_{g}}$ to $\mathcal{C}$ with Superglue \cite{c37}.  The highest density of the matched keypoints within each bounding box in $\mathcal{Z}_r$ determines the most likely corresponding object. The matched object is denoted by:

\begin{align}
   \overline{\mathcal{Z}_{{r}}}  = \mathop{\arg\max}\limits_{\beta_{r_i} \in \mathcal{Z}_r} \sum   _{\mathbf{k}_j \in \overline{\beta_{g}}} \mathbbold{1}(\mathbf{k}_j \in \beta_{r_i})
   \vspace{-5pt}
\end{align}

\begin{figure*}[t]
      \centering
      \includegraphics[width=0.7\textwidth]{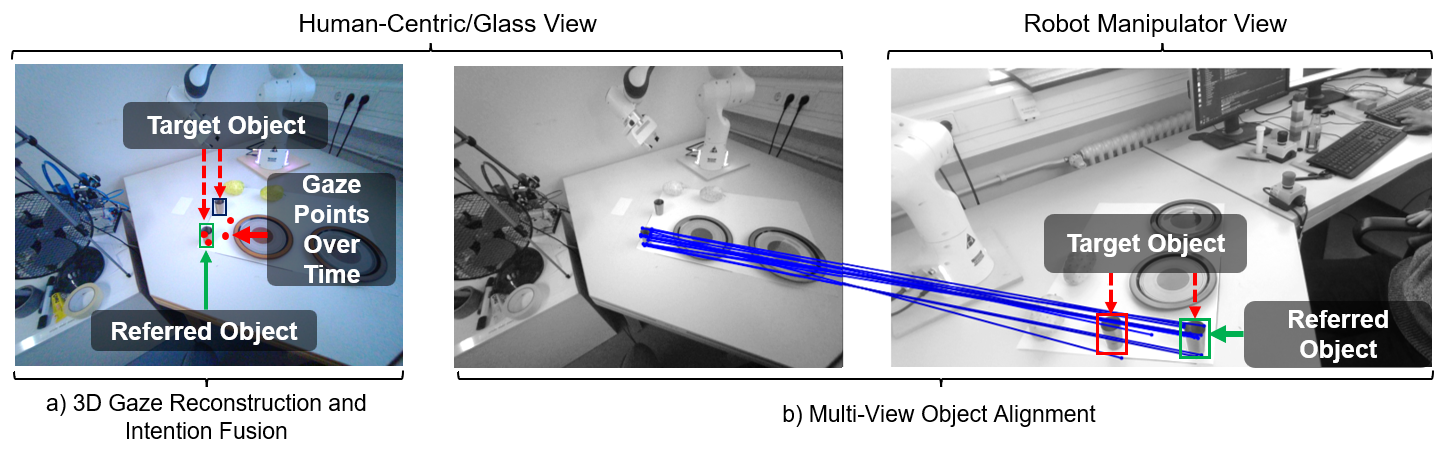}
    \vspace{-10pt}
      \caption{ Human view intention fusion and multi-view intention alignment.
      The user's speech and gaze are aligned using a LLM-Agent, and the user intention is then processed by a multi-view alignment module to get the referred object within the robot manipulator view.}
      \label{figurelabel4}
 \vspace{-15pt}
\end{figure*}

Where $\mathbbold{1}(\cdot)$ is an indicator function that equals one if the matched keypoint $\mathbf{k}_j$ falls within the segmentation mask $\beta_{r_i}$. This ensures robust object alignment across the two viewpoints. The general process of human intention fusion and multi-view intention alignment is illustrated in Fig. \ref{figurelabel4}.

\subsection{Planning Policy Generation}
\label{policy}
The final stage of our proposed system involves generating a planning policy to execute the human intention. In our proposed system, the state space $X$ in Eq. \ref{eqpolicy} is defined as $X =\left \{  \overline{\mathcal{Z}_{r}}, \ \mathbf{O}_1, \ \mathcal{S} \right \} $ including referred object, auxiliary from the first LLM-Agent and user's command.

To translate this state space into an executable sequence  
of parameterized action primitives, we employ a  
LLM-based planning policy generator.  
We adapt GPT-4o with frozen checkpoints from 2024-08-06,  
using a temperature of 1 and input $X$ for policy generation.  
The model is prompted to generate semantic-level  
parameterized action primitives in JSON format, denoted as $\pi$. In particular, our action primitives include both composite actions and atomic actions to flexibly support a wide range of task compositions:
\begin{enumerate}
    \item \textbf{Composite actions} involve task-oriented behaviors commonly encountered in daily HRI scenarios:
    \begin{enumerate}
        \item Pick: grasp an object and lift it from the surface.
        \item Put: place an object at a designated location.
        \item Pour: tilt and empty the contents of a cup into a target container.
        \item Swap: exchange the positions of two objects.
        \item MoveTo: move the end-effector to a position.
    \end{enumerate}
    \item \textbf{Atomic actions} offer fine-grained control over the robot’s motion and end-effector state:
    \begin{enumerate}
        \item MoveX, MoveY, MoveZ: move the end-effector along the x, y, or z axes.
        \item OpenGripper: open the gripper.
        \item CloseGripper: close the gripper.
        \item Rotate: rotate the end-effector.
    \end{enumerate}
\end{enumerate}

To generate a reasonable policy, we design structured prompts that enforce strict constraints on parameterized action primitives usage, the coordinate system, and the expected output format. By providing task descriptions, parameterized action primitive definitions, example tasks, and structured planning steps, we ensure that the LLM outputs consistent, interpretable, and physically feasible actions, reducing ambiguity and enhancing execution reliability. \textcolor{black}{The prompt is shown in Appendix. \ref{PromptB}.}  An example input and output of the planning policy generation is shown as follows:

\vspace{10pt}
\noindent\begin{minipage}{\linewidth}
\raggedright
\textbf{Input:} \\
\textbf{$\mathbf{O}_1$:} \texttt{$(object, \ position),(apple, \  table),(\Delta t_{apple}, \ \Delta t_{here})$} \\
\textbf{$\overline{\mathcal{Z}_r}$:} \texttt{$(Position_{apple}, \ Position_{here})$} \\
\textbf{$\mathcal{S}$:} \texttt{$please \ put \ the \ apple \ here \ on \ the \ table$} \\
\textbf{Output:} \\
\textbf{$\pi$:} \texttt{$[open \ gripper], \ [pick, \ ['apple', \ position_{apple}]]$ \\ 
\ \  $,\ [close \ gripper], \  [put,  \ ['table', \ position_{there}]]$ \\
\ \  $  [open \ gripper]$} 
\end{minipage}

\textcolor{black}{To mitigate potential LLM hallucinations or infeasible outputs, we further design a logic-checking module based on prior works \cite{c17, c39} that validates the sequence against basic executability and safety constraints (e.g., the gripper must be opened before a pick; consecutive pick actions are not allowed without an intervening put; required arguments/objects must exist). If the generated plan fails these checks, it is rejected (and the system requests regeneration), preventing hallucinated or logically invalid action sequences from being executed.}

\subsection{\textcolor{black}{Deterministic Conflict Resolution Strategy}}


\textcolor{black}{We consider four scenarios: $C_1$) speech and gaze indicate the same object; $C_2$) gaze occurs outside the time window or is missing; $C_3$) a large spatial discrepancy remains between gaze points and the object mask; and $C_4$) the gazed object is not the object implicitly intended by the user.}


\textcolor{black}{Based on Sec. \ref{Human View Intention Alignment}, gaze samples and the speech-derived time window $\Delta t$ from the LLM agent are fused. An intended referred object is selected if $\overline{\mathcal{Z}_{g}} $ (referred from Eq. \ref{zg}) lies within a fixed distance (80 pixels, approximately 5cm on the desk) from the boundary of the object mask ($C_1$). When a valid object is identified, the robot sequentially moves its end-effector above the object and explicitly queries the user with a yes/no interactive confirmation before execution, in order to avoid cases where the user looks at one object while intending another ($C_4$). If no valid object is found within $\Delta t$ ($C_2$), the $\Delta t$ is relaxed to include neighboring words ($\pm 1$). If ambiguity remains ($C_3$), the robot sequentially moves its end-effector above candidate objects (all objects within $c_{target}$) and queries the user with a yes/no interaction confirmation before execution. If there are still no executable tasks, the robot guides the user and waits for the new command. The demonstration is shown in Appendix. \ref{demon} Fig. \ref{figdemon}}.

\textcolor{black}{
For safety, following prior work \cite{c17}, collision avoidance and constrained motion execution are handled using established sweep-volume–based methods, and the LLM agent is prompted to raise the end-effector above the tallest detected object to maintain sufficient clearance. In ambiguous or failure cases, the system either requests user clarification or safely stops, ensuring predictable and safe robot behavior. }

\section{Experiment Setup}
\label{experimentsetup}
\subsection{Perception and Manipulator Setup}

\begin{figure}[t]
      \centering
      \includegraphics[width=0.35\textwidth]{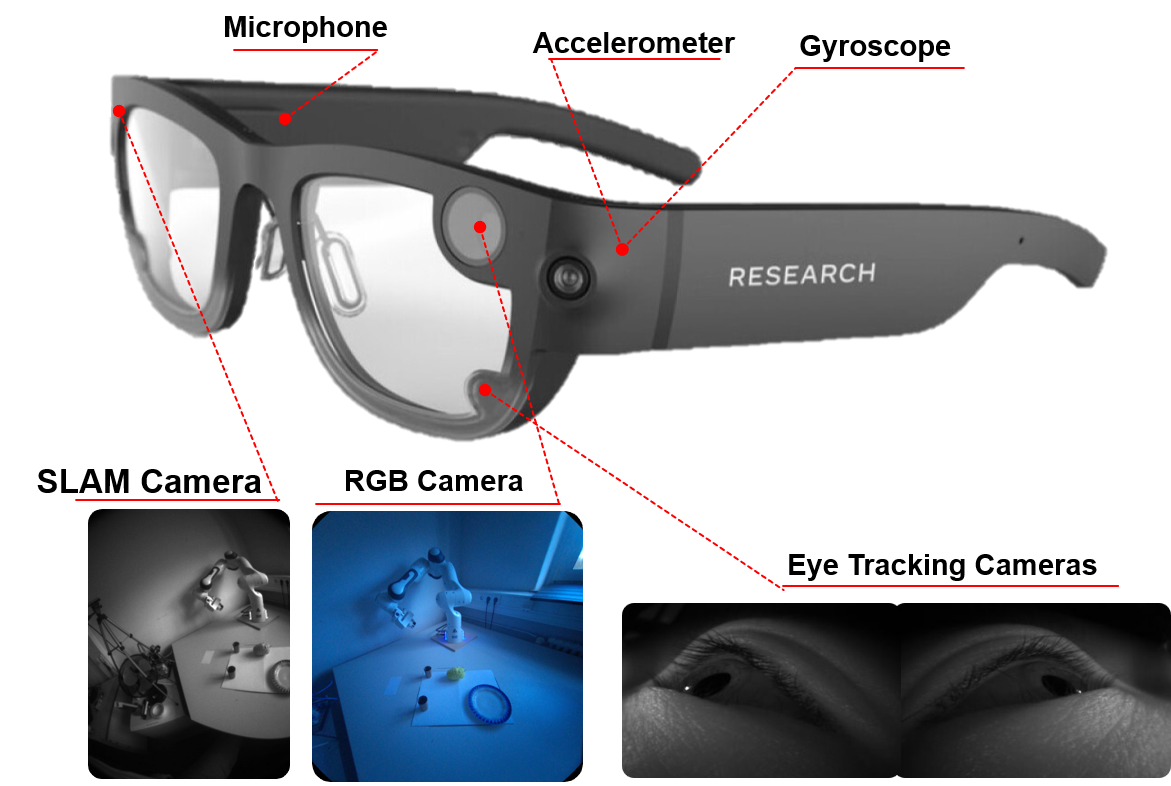}
     
      \caption{\textcolor{black}{Meta ARIA Glasses: Sensor Setup.}}
      \label{ARIA}
     \vspace{-20pt}
\end{figure} %

\begin{figure*}[t]
      \centering
      \includegraphics[width=0.85\textwidth]{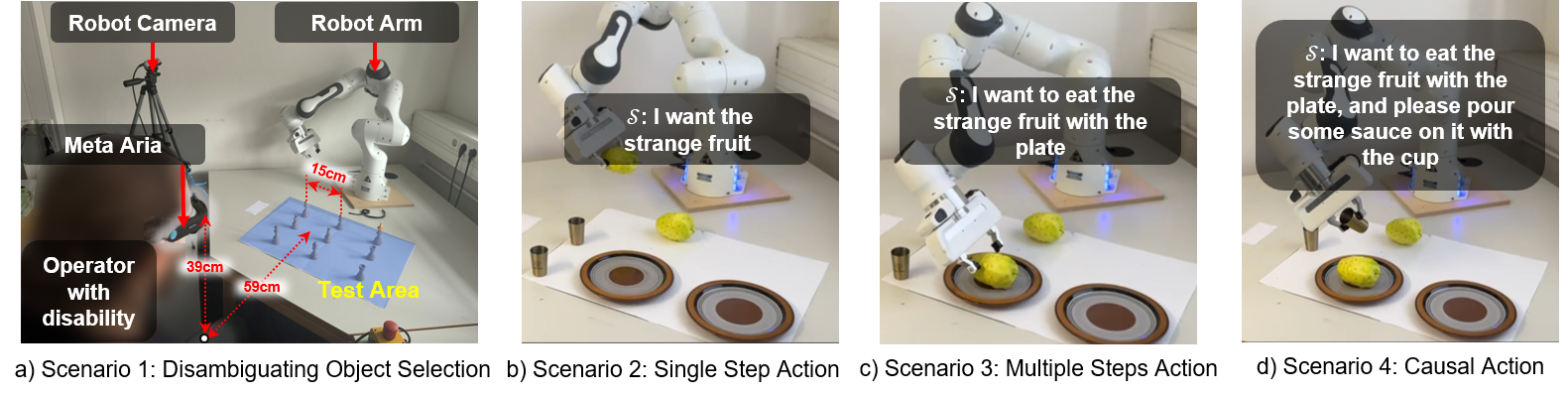}
      \vspace{-5pt}
      \caption{ \textcolor{black}{Overview of Experimental Scenarios. Scenario 1 evaluates object selection among similar items. Scenarios 2-4 assess real-world robotic task execution at increasing complexity.}}
      \label{figure6}
 \vspace{-10pt}
\end{figure*}

Visual inputs for human view are collected through the Meta ARIA glasses\footnote{\url{https://www.projectARIA.com/}}, which also capture auditory signals through the built-in microphone. Examples of the outputs of the sensors used in the experiment and the reference frames are shown in Fig. \ref{ARIA}. The robot view is collected through the Intel Realsense D435i RGBD camera.
Our system has been tested with individuals in real-world environments, as illustrated in Fig. \ref{figure6} (a-d). 
The scene includes a Franka Emika Panda 7DOF robot manipulator  
and various manipulation objects.  
An Intel D435i RGBD camera is positioned to monitor the workspace,  
ensuring clear visibility of both the environment and the robot.  
Multimodal data processing is performed on a computer equipped  
with an RTX 4080 GPU, utilizing 6105 MB of GPU memory during interaction. 

\subsection{Description of Experimental Scenarios}
\label{v-b}
To evaluate our system, we designed four experimental scenarios as shown in Fig. \ref{figure6}. \textcolor{black}{The detailed evaluation of the tasks is shown in Appendix. \ref{Evaluation of Tabletop Task}.}
\begin{enumerate}
\item \textbf{Object Selection Among Similar Items:}
In scenario 1 ($S_1$) shown in Fig. \ref{figure6} (a), the test area contains nine visually similar pawns, each with a distance of 15~cm between them, requiring the user to select one of the pawns designated by the experiment organizer by gazing at the bottom center of the pawns. The gaze error was reported in this scenario. \textcolor{black}{Each participant makes five trials for each baseline approach and our proposed approach, with the approach used in each trial being completely randomized.}
    \item \textbf{Task Execution in a Complex Environment:}
As shown in Fig.~\ref{figure6} (b-d),  
Scenarios 2-4 ($S_2$-$S_4$) evaluate the execution of robotic tasks
in increasingly complex real-world environments.  
The test area contains two similar cups,  
two similar exotic fruits (unknown name), and two similar plates.  
We categorize tasks into three levels of complexity:  1) \textbf{Single-Step Action ($S_2$):} Tasks requiring a single, direct action and where only one target object is referred to, such as picking up an object. 2) \textbf{Multi-Step Action ($S_3$):} Sequential actions requiring multiple steps to complete a task and where two target objects are referred to, such as putting an object on a plate. 3) \textbf{Causal Action ($S_4$):} Tasks where an earlier action influences a later one,  
   involving more than two target objects.  
   For example, placing a fruit on a plate,  
   then pouring liquid from a cup onto it  
   without explicitly mentioning the plate's position again. \textcolor{black}{Each participant makes five trials for each baseline approach and our proposed approach, with the approach used in each trial being completely randomized.}
   \item \textbf{\textcolor{black}{Deterministic Resolution Strategy: }}\textcolor{black}{Each participant completed three experimental blocks under the $S_2$ interaction scenario. 
In the first block, the deterministic resolution strategy was disabled, and the participant's command and true intention were consistent; Each participant performed five trials under this setting. 
In the second block, the deterministic resolution strategy remained disabled, but the participant's command and true intention were inconsistent; Each participant again performed five trials. 
In the third block, the deterministic resolution strategy was enabled, and the command and user intention were inconsistent; Each participants performed an additional five trials.}

\end{enumerate}

Moreover, a user study was conducted to evaluate user satisfaction and experience. For experiments and user study, 
a total of \textcolor{black}{42} participants were recruited, comprising 18 participants from Germany and \textcolor{black}{24} from Singapore. \textcolor{black}{Among the Singaporean participants, \textcolor{black}{20} have mild upper-limb impairments. Furthermore, we formulated hypotheses regarding user perceptions (\textbf{H1}, \textbf{H2}), and performed hypothesis testing to evaluate statistical significance. The experimental setup for user study and the distribution of the participants is shown in Appendix. \ref{userstudy} and Fig. \ref{figuredisable}.}


\subsection{Evaluation Metrics} 
To quantitatively assess system performance, we use two key metrics: \textbf{success rate} and \textbf{interaction time}. The success rate is defined as the proportion of correctly executed tasks ($N_{\text{correct}}$) to the total number of trials ($N_{\text{total}}$), given by $(N_{\text{correct}} / N_{\text{total}})* 100\%$.
Interaction time represents the efficiency of the system, measured as the time from when the user begins issuing a command to when the system completes interaction. This metric eliminates variations caused by different robot platforms, network situations, and computing configurations, ensuring a fair comparison across methods.

We have also included a supplementary video that presents experimental results, illustrating the system’s performance across various tasks and environments.


\section{Results and Discussion}

\subsection{Baseline Selection}

In our proposed FAM-HRI, we compare our FAM-HRI with the various SOTA interaction methods utilizing different modalities. CaP \cite{c38} and ProgPrompt \cite{c39} are language-only interactions, making $S_1$ inefficient, as users would be required to specify the exact row and column of the pawns. 
GesSentence \cite{c5} utilizes gesture-only interaction, HapGloves \cite{c41} employs wearable tactile gloves for users to teleoperate the robot, while PhysicalDemo \cite{c42} leverages impedance-based feedback, where the user physically guides the robot through specific tasks. NVP-HRI \cite{c17} combines language and deictic posture for interaction, while GIRAF \cite{c43} fuses language with deictic gestures to achieve efficient object selection. These approaches are efficient in selecting similar objects. However, they rely on body movements and are less convenient for users with mobility or motor disabilities. FreeView \cite{c24} enables robot teleoperation through a virtual GUI controlled by the user's gaze. Additionally, since Gesture Sentence uses pre-trained supervised networks for environment observation, therefore cannot deal with unfamiliar objects in the scene. Comparison results are shown in Tab. \ref{table_comparison}. \textcolor{black}{The detailed user commands are shown in Tab. \ref{tab:task}.}  



\begin{table*}[thp]
\caption{\textcolor{black}{Comparison of different HRI approaches. Best results are bolded to support our contribution claims. }}
\vspace{-10pt}
\label{table_comparison}

\begin{center}
\begin{tabular}{ccccccccccccccc}
\hline
\hline
\multicolumn{1}{c}{\multirow{2}{*}{Method}} & \multicolumn{1}{c}{\multirow{2}{*}{\shortstack{\\Interaction\\Modality}}} & \multicolumn{1}{c}{\multirow{2}{*}{\shortstack{\\Hands\\ Free?}}}&   \multicolumn{1}{c}{\multirow{2}{*}{\shortstack{\\Voice\\ Free?}}}&        \multicolumn{1}{c}{\multirow{2}{*}{\shortstack{\\Similar\\ Object?}}}&  \multicolumn{1}{c}{\multirow{2}{*}{\shortstack{\\Unfamiliar\\ Object?}}}&         \multicolumn{4}{c}{\shortstack{\\Success Rate $\uparrow$ (\%)}}& \multicolumn{4}{c}{\shortstack{\\Interaction Time $\downarrow$ (s)}}   \\ \cmidrule(r){7-10} \cmidrule(r){11-14} 
 & &  &   & &  &   $S_1$ & $S_2$ &  $S_3$ & $S_4$ &  $S_1$&   $S_2$& $S_3$&$S_4$\\ \hline
 \shortstack{\\ProgPrompt \cite{c39} } &  \shortstack{\\Language} &\textcolor{green} {\ding{51}}&\textcolor{red}{\ding{55}}&\textcolor{red}{\ding{55}}&\textcolor{green}{\ding{51}}&12&73&52&16&4.7&2.4&6.4&9.7\\ \hline
  \shortstack{\\CaP \cite{c38}}  &\shortstack{\\Language} &\textcolor{green}{\ding{51}}&\textcolor{red}{\ding{55}}&\textcolor{red}{\ding{55}}&\textcolor{green}{\ding{51}}&36&97&$74$&61&4.7&3.7&6.9&11.5\\ \hline
  \shortstack{\\GesSentence \cite{c5}} &\shortstack{\\ Gesture}&\textcolor{red}{\ding{55}}&\textcolor{green}{\ding{51}}&\textcolor{green}{\ding{51}}&\textcolor{red}{\ding{55}} & 67&82&73&68&5.4&5.1&10.2&22.8  \\  \hline
 \shortstack{\\HapGloves \cite{c41}}& \shortstack{\\ Haptic(teleoperation)}&\textcolor{red}{\ding{55}}&\textcolor{green}{\ding{51}}&\textcolor{green}{\ding{51}}&\textcolor{green}{\ding{51}}&91&98&$\underline{96}$&89&13.9&10.8&18.7&33.1 \\ \hline
\shortstack{\\PhysicalDemo \cite{c42}} & Impedance&\textcolor{red}{\ding{55}}&\textcolor{green}{\ding{51}}&\textcolor{green}{\ding{51}}&\textcolor{green}{\ding{51}}&\textbf{100}&\textbf{100}&61&52&6.3&5.8&12.3&31.7 \\\hline
\shortstack{\\FreeView \cite{c24}} & Gaze(teleoperation)&\textcolor{green}{\ding{51}}&\textcolor{green}{\ding{51}}&\textcolor{green}{\ding{51}}&\textcolor{green}{\ding{51}}& 94 &\textbf{100}&\textbf{100}&\textbf{100}&27.8&63.4&106.3&192.7 \\\hline
\shortstack{\\NVP-HRI \cite{c17}} & \shortstack{\\ MultiModal}&\textcolor{red}{\ding{55}}&\textcolor{red}{\ding{55}}&\textcolor{green}{\ding{51}}&\textcolor{green}{\ding{51}}&88&\textbf{100}&93&86&$2.3$&2.2&4.6&11.2 \\\hline
\shortstack{\\GIRAF \cite{c43} } &  \shortstack{\\MultiModal}&\textcolor{red}{\ding{55}}&\textcolor{red}{\ding{55}}&\textcolor{green}{\ding{51}}&\textcolor{green}{\ding{51}}&85&\textbf{100}&84 &77&$\underline{2.1}$&$\underline{1.6}$&$\underline{2.9}$&$\underline{6.8}$\\\hline
\shortstack{\\FAM-HRI (Ours)} &  \shortstack{\\ MultiModal}&\textcolor{green}{\ding{51}}&\textcolor{red}{\ding{55}}&\textcolor{green}{\ding{51}}&\textcolor{green}{\ding{51}}&$\underline{96} $&\textbf{100}&94&$\underline{92} $&\textbf{1.8}&\textbf{1.3}&\textbf{2.2}&\textbf{5.9}\\
\hline
\hline
\end{tabular}  
\end{center}
\footnotesize{ \textbf{Remark:} 
\textcolor{black}{All experiments were conducted under identical conditions, with identical tasks, identical hardware, and identical evaluation protocols. We collect data through international cooperation.}
}
\vspace{-10pt}
\end{table*}

\subsection{Object Selection Among Similar Items and Gaze Error}

In $S_1$, as described in Sec. \ref{v-b} and illustrated in Fig. \ref{figure6} (a), all \textcolor{black}{42} participants received verbal instructions on how to use FAM-HRI \textcolor{black}{and other baseline approaches}. To evaluate our FAM-HRI in distinguishing between multiple similar objects, each participant completed five trials \textcolor{black}{using all baseline approaches and our proposed method} of pawn selection as instructed. 
The results, presented in Tab. \ref{table_comparison}, show that our system achieved the second-highest success rate among all baseline methods in $S_1$, only lower than Physical Demo \cite{c42}, which requires the user to physically guide the robotic arm near the target object. However, FAM-HRI significantly outperformed gesture-only and language-only methods, demonstrating its reliability in challenging and cluttered environments. We obtained the gaze error by measuring the distance between the gaze point and the center of the bottom of the pawn. With each participant wearing ARIA glasses correctly, the gaze error is $1.58\pm0.62$ cm.

\subsection{Task Execution in Complex Environment}

In $S_2-S_4$, described in Sec. \ref{v-b} and illustrated in Fig. \ref{figure6} (b-d), the same \textcolor{black}{42} participants were asked to complete three levels of tasks using objects in the scene ($S_2$: Single-Step Action, $S_3$: Multi-Step Action, $S_4$: Causal Action). Each participant performed five trials \textcolor{black}{using all baseline approaches and our proposed method} per task level.

As shown in Tab. \ref{table_comparison}, FAM-HRI outperformed all baseline methods (except for the two teleoperation baselines, FreeView and HapGloves), achieving the highest success rate across all task levels, demonstrating its robustness in complex real-world scenarios. Our success rate exceeds that of GIRAF \cite{c43} and NVP-HRI \cite{c17}, both of which rely on an RGB-D camera to capture the 3D human skeleton, which is prone to sensor noise and ambient occlusion that reduces the overall success rate of the task execution.

\subsection{Comparison of Effectiveness}
The results in Tab.~\ref{table_comparison} present the average interaction time across participants  
for each of the four scenarios.  
FAM-HRI achieves the shortest interaction time  
among baseline methods, demonstrating its efficiency in processing seamless multimodal inputs.  
Unlike language-only methods, which require precise verbal descriptions,  
or gesture-based approaches, which depend on physical movement,  
FAM-HRI leverages gaze-language fusion to streamline intent inference.  
Furthermore, methods relying on teleoperation or physical demonstration  
exhibit longer interaction times due to manual corrections  
and additional user input steps.  
These results confirm that FAM-HRI enables faster interactions,  
making it a practical and efficient solution for real-world  
human-robot collaboration.

\subsection{Qualitative Field Survey on User Experience}

To rigorously assess the proposed method, we conducted a user survey to evaluate interaction experience, including participants with mild upper-limb impairments.
All participants were asked to complete a survey at the end of the experiment.
Participants responded to questions using a five-point Likert scale, 
assessing FAM-HRI and baseline methods in terms of
object selection accuracy, device flexibility, approach modernity,  
ease of use, and interaction efficiency. As shown in Tab.~\ref{experience},  
users rated FAM-HRI higher than all baselines across qualitative measures.  
Notably, it outperformed baselines in efficiency and accuracy,  
with all participants favoring FAM-HRI for its seamless  
integration of gaze and language signals.  
Additionally, most users found our system more modern  
and easier to operate, emphasizing its intuitive design. 

This user study investigated two main hypotheses:
\begin{enumerate}
    \item \textbf{H1:} FAM-HRI is preferred by participants compared to all other baseline methods.
    \item \textbf{H2:} FAM-HRI is quantitatively more effective and accurate compared to all other baseline methods.
\end{enumerate}

\textbf{H1: } Participants rated FAM-HRI higher than all baseline methods as ``light and flexible" ($p<0.001$), ``modern" ($p<0.001$), and ``simple to use" ($p<0.001$).

\textbf{H2: } Participants further rated the execution of their personal preferences, FAM-HRI, as ``accurate" ($p<0.001$) and ``efficient" ($p<0.001$).

\textcolor{black}{To examine whether user background influenced subjective evaluation, we conducted a one-way Analysis of variance
(ANOVA) to compare survey ratings between normal (N) and disabled (D) participants across all five evaluation dimensions. 
The analysis (as shown in Tab. \ref{tabanova}) revealed no statistically significant differences between the two groups under all dimensions ($p > 0.05$) except ``ease of use" ($p=0.003$). More participants with upper limb disabilities found our method easier to use. This indicates that both normal and disabled participants perceived the system similarly in terms of all five dimensions.}

The results show that our proposed approach
significantly reduced physical effort,  
offering a more accessible and user-friendly solution  
for human-robot collaboration.

\begin{table}[t]

\caption{\textcolor{black}{Results of User Experience Survey.}}
\vspace{-10pt}
\label{experience}
\begin{center}
\setlength{\tabcolsep}{4pt}
\begin{tabular}{cccccccccccc}
\hline
\hline
\multicolumn{1}{c}{\multirow{2}{*}{Rate}} &  \multicolumn{2}{c}{\shortstack{\\Accurate }} &  \multicolumn{2}{c}{\shortstack{\\Flexible }}&  \multicolumn{2}{c}{\shortstack{\\Modern }}&  \multicolumn{2}{c}{\shortstack{\\Simple }} &  \multicolumn{2}{c}{\shortstack{\\Efficient }} \\ \cmidrule(r){2-3} \cmidrule(r){4-5} \cmidrule(r){6-7} \cmidrule(r){8-9} \cmidrule(r){10-11}
 &   N & D &   N & D&   N & D&   N & D&   N & D\\ \hline
  Strongly Agree & 5& 8&2&4&9&12&9&18&14&13  \\ \hline
 Agree & 10&6&11&9&5&3&6&2&6&7  \\ \hline
Neutral & 3&2&4&4&2&1&3&2&2&0  \\ \hline
Disagree & 4&3&3&2&4&4&3&0&0&0  \\ \hline
Strongly Disagree &  0&1&2&1&2&0&2&0&0&0 \\ 
\hline 
\hline
\end{tabular}  
\end{center}
\footnotesize{\textcolor{black}{\textbf{Remark:} Participants evaluated the proposed system along five dimensions: \emph{Accurate}, \emph{Flexible}, \emph{Modern}, \emph{Simple}, and \emph{Efficient}.
The numbers indicate the count of participants selecting each rating. 
N denotes normal participants, and D denotes participants with mild upper-limb disabilities.}

}

\end{table}

\begin{figure}[t]
      \centering
      \includegraphics[width=0.45\textwidth]{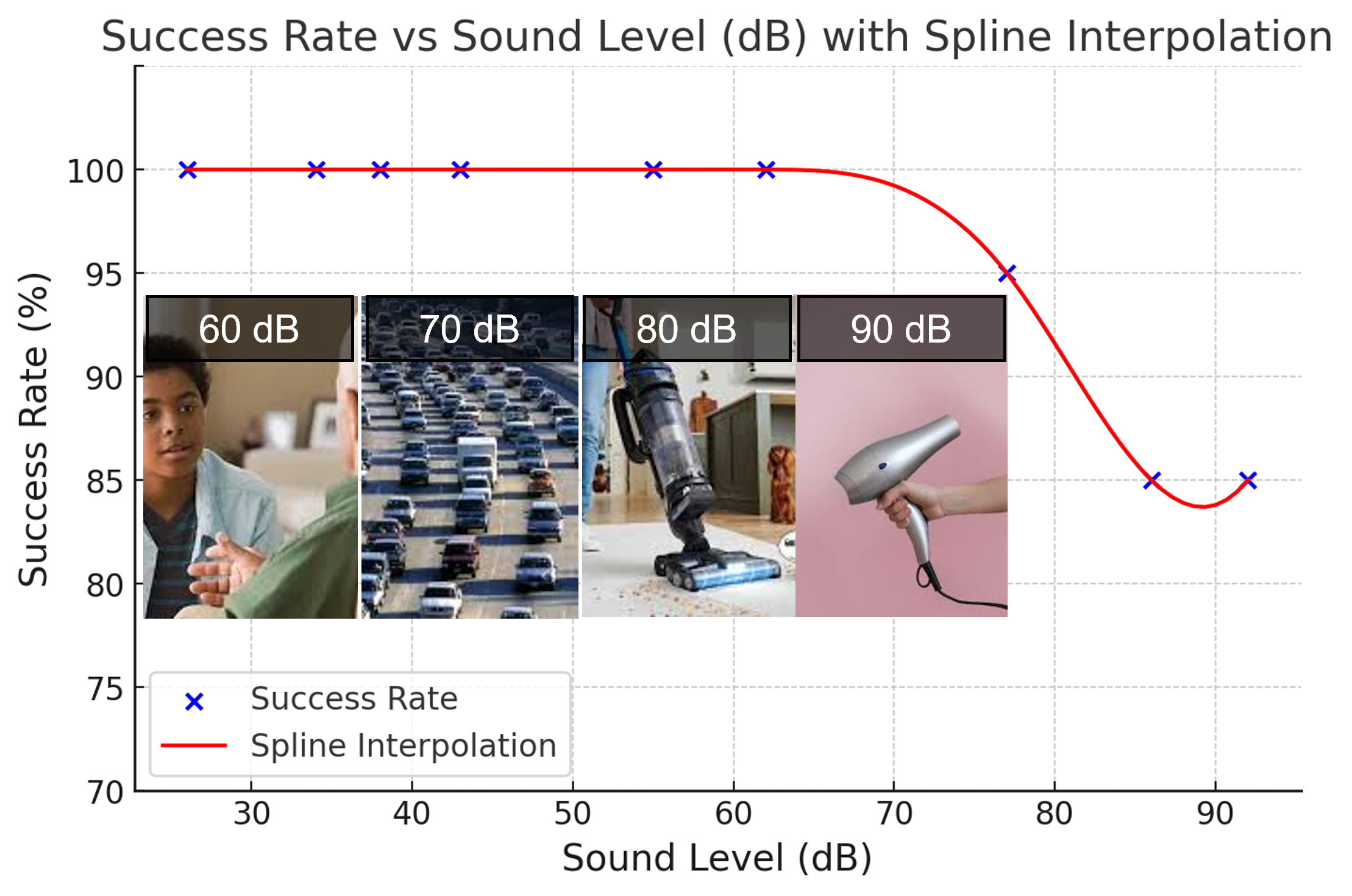}
      \vspace{-10pt}
      \caption{System robustness evaluation under varying background noise. The system consistently achieved 100\% success rate up to 70dB.}
      \vspace{-10pt}
      \label{noise}
\end{figure} %

\subsection{Task Execution under varying noise levels}
To evaluate the robustness of FAM-HRI in real-world conditions, we conducted experiments under varying ambient noise levels. 
The noise intensity ranged from 20 dB to 90 dB, covering common noise conditions, such as conversation, traffic, vacuum cleaner, and hairdryer. According to the Level Comparison Chart\footnote{\url{http://ehs.yale.edu/sites/default/files/files/decibel-level-chart.pdf}}, hearing loss may result from sustained exposure to 90 dB or more. The setup for the noisy environment is shown in Appendix \ref{Noise Environment} 

The results, presented in Fig. \ref{noise}, illustrate that our proposed FAM-HRI maintains a high success rate even under increasing noise levels. Here, the user needs to complete a pick-placing task. The system consistently achieved 100\% success rate up to 70 dB, indicating strong resilience to moderate noise conditions. Even at 90+ dB, where typical speech recognition systems often fail, FAM-HRI still maintained an 85\% success rate, demonstrating its robustness in handling speech input under challenging conditions. 

Our two LLM-Agent frameworks further enhance robustness against background noise. The LLM-Agent for human view command processing incrementally analyzes task-relevant verbs, nouns, and pronouns in the user’s command, filtering out irrelevant words before passing a cleaned and structured instruction to the agent for policy generation. 


\subsection{LLM Inference}
\subsubsection{\textcolor{black}{Latency Analysis}}
Tab. \ref{table_latency} presents the inference latency of the two LLM-Agents under different experimental scenarios. The reported values include the average response time for human view command processing and planning policy generation, measured across all user trials. 

The results show that command processing latency varies between 1353ms and 2202ms, with increased latency in more complex scenarios due to additional language reasoning. Similarly, the policy generation latency ranges from 1281ms to 1997ms, reflecting the increasing complexity of task planning as the number of required action primitives grows.

Despite these variations, our proposed FAM-HRI maintains a reasonable inference latency, demonstrating its feasibility for human-robot collaboration.

\begin{table}[t]
\caption{\textcolor{black}{Inference latency under different conditions}}
\vspace{-10pt}
\label{table_latency}
\begin{center}
    \centering
    \begin{tabular}{ccc}
     \hline
      \hline
     \multicolumn{1}{c}{\multirow{2}{*}{\shortstack{\\Scenario}}}  &  \multicolumn{1}{c}{\multirow{2}{*}{\shortstack{\\Latency of  Human View \\ Command Processing}}} & \multicolumn{1}{c}{\multirow{2}{*}{\shortstack{\\ Latency of \\ Policy Generation }}}   \\
     \\
     & & \\ \hline
    $S_1$     & $1353\pm753 $ms&$1281\pm692$ms  \\ \hline
    $S_2$     & $1624\pm660$ms & $1535\pm946$ms\\ \hline
    $S_3$     & $1855\pm725$ms& $1803\pm891$ms\\ \hline
    $S_4$     &$2202\pm936$ms &$1997\pm 861$ms \\
    \hline
    \hline
    \end{tabular}
    \end{center}
    \footnotesize{\textbf{Remark:} The reported latency values correspond to inference using a full-scale LLM \textbf{without} any quantization \cite{hu2022lora}, distillation \cite{bucila2006model}, or retrieval-augmented generation (RAG) \cite{lewis2020retrieval}. Incorporating such optimization strategies is expected to yield substantial reductions in latency. }
\end{table}

\subsubsection{\textcolor{black}{Baseline comparison}}  

\textcolor{black}{To evaluate the impact of different LLMs on system performance and reproducibility, we conducted a baseline comparison between proprietary and open-source LLM agents under the $S_3$ scenario.}

\textcolor{black}{All open-source LLaMA \cite{DBLP:journals/corr/abs-2302-13971} models were deployed on two NVIDIA A100 GPUs, while GPT-4o was accessed via an external API. As reported in Tab~\ref{table_LLM agents}, 
although GPT-4o achieves the highest success rate overall, the performance gap between GPT-4o and LLaMA~3.3~70B remains moderate. Importantly, the strong performance of open-source models demonstrates that the proposed framework is not inherently dependent on a specific proprietary LLM, and can be instantiated with open-source alternatives while maintaining high effectiveness. This significantly improves the reproducibility and extensibility of the system.}

\begin{table}[t]
\caption{\textcolor{black}{Baseline comparison of LLM agents}}
\vspace{-10pt}
\label{table_LLM agents}
\begin{center}
    \centering
    \begin{tabular}{ccc}
     \hline
      \hline
     Methods  &  Success Rate $\uparrow$  & Average latency 
     \\ \hline
     Llama3.3 70B
    & 0.91&$2938\pm192$ms  \\ \hline
    Llama3.1 70B
    & 0.88&$6163\pm304$ms  \\ \hline
   Llama3.1 8B
   & 0.63 & $6389\pm417$ms\\ \hline
    Gpt4o t=0.7
    & \textbf{0.94}& $3579\pm812$ms\\ \hline
    Gpt4o t=1
    &0.92 &$3658\pm 783$ms \\ 
    \hline
    \hline
    \end{tabular}
    \end{center}
    \footnotesize{\textcolor{black}{\textbf{Remark:} The experiment was conducted in the $S_3$ (Multi-Step Action).}} 
    \vspace{-15pt}
\end{table}

\subsection{\textcolor{black}{Evaluation of Deterministic Conflict Resolution Strategy}}
    \begin{figure}[t]
      \centering
      \includegraphics[width=0.45\textwidth]{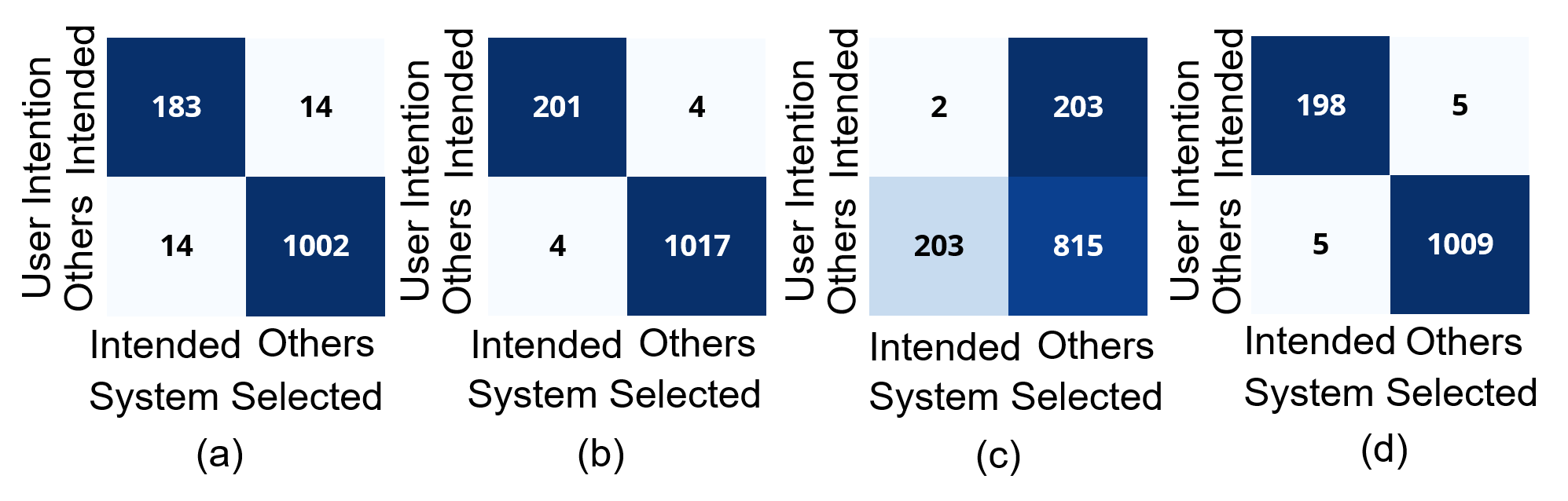}
     \vspace{-10pt}
      \caption{\textcolor{black}{Confusion matrix analysis of intention recovery under different interaction conditions. (a) Command and user intention are consistent, without the proposed conflict resolution strategy; (b) Command and user intention are consistent, with the proposed conflict resolution strategy; (c) Command and user intention are inconsistent, without the proposed conflict resolution strategy; and (d) Command and user intention are inconsistent, with the proposed conflict resolution strategy.}}
      \label{confusionmatrix}
      \vspace{-10pt}
\end{figure} %

\textcolor{black}{We evaluate the proposed deterministic conflict resolution strategy using confusion matrices defined over \emph{user intention} versus \emph{system selection}. Specifically, the confusion matrix rows correspond to the object that the user intended to refer to, while the columns correspond to the object ultimately selected by our proposed system. }

\textcolor{black}{We distinguish two groups of interaction conditions. In $C_1$-$C_3$, the user’s intended referred object is consistent with the object mentioned in the multimodal command, while gaze may be consistent ($C_1$), temporally misaligned ($C_2$), or otherwise spatially misaligned ($C_3$). In contrast, $C_4$ represents the more challenging case where the user’s true intention differs from the object explicitly mentioned in the command (e.g., the user looks at one object while intending another), which can lead to unintended execution without explicit disambiguation.}

\textcolor{black}{In Fig. \ref{confusionmatrix}, confusion matrices illustrating the relationship between user intention and system-selected object under four interaction conditions. The confusion matrices aggregate all non-intended target objects into a single “Other” class, emphasizing the system’s ability to recover the user’s true intention. The results show that the proposed conflict resolution strategy significantly reduces mis-selections, particularly in cases where command and user intention are inconsistent.}

\vspace{-10pt}
\subsection{Ablation studies}

\subsubsection{\textcolor{black}{Gaze Trajectory Filter Function}} 

\begin{table}[t]
\caption{\textcolor{black}{Ablation study of gaze trajectory filter function.}}
\vspace{-10pt}
\label{ablation alignments}
\begin{center}
\setlength{\tabcolsep}{2pt}
\begin{tabular}{ccccccccc}
\hline
\hline
\multicolumn{1}{c}{\multirow{3}{*}{Approach}} &  \multicolumn{8}{c}{\shortstack{\\Correct Referred Probability $\uparrow$ }}  \\ \cmidrule(r){2-9} 
&  \multicolumn{4}{c}{\shortstack{\\Short interval ($N \le  3$)}} &  \multicolumn{4}{c}{\shortstack{\\Long interval ($N > 3$)}} \\  \cmidrule(r){2-5}   \cmidrule(r){6-9} 
 &    $G_1$ & $G_2$ &  $G_3$ & $G_4$ &    $G_1$ & $G_2$ &  $G_3$ & $G_4$ \\ \hline
Uniform weight & 0.89   &0.73&0.72&0.21& \textbf{0.98} &0.94&0.92&0.13 \\\hline
Linear weight & 0.92 &\textbf{0.94}& 0.88 &0.95& \textbf{0.98} &0.95&0.77&0.84 \\\hline
OPTICS clustering& \textbf{0.96} &-&-&-&\textbf{0.98} &\textbf{0.97}&0.95&0.03 \\\hline
Last gaze & 0.93 &0.92&0.58&0.89& 0.94 &0.93&0.35&0.72 \\\hline
Our w/o OPTICS & \textbf{0.96} &\textbf{0.94}&\textbf{0.91}&\textbf{0.98}&\textbf{0.98} &\textbf{0.97}&0.79  &\textbf{0.93} \\\hline
Our & \textbf{0.96} &\textbf{0.94}&\textbf{0.91}&\textbf{0.98}& \textbf{0.98} &\textbf{0.97}&\textbf{0.95}&\textbf{0.93 }\\
\hline
\hline
\end{tabular}   
\end{center}
\footnotesize{ \textcolor{black}{\textbf{Remark:} Correct Referred Probability = correct selected referred object / number of trials, ``-" indicate that OPTICS clustering over short intervals does not work. In our implementation for linear weighting, the last gaze point receives a weight of 0.49 while the first receives 0.1, followed by normalization.}

}

\end{table}

\textcolor{black}{Tab~\ref{ablation alignments} reports an ablation study evaluating different gaze trajectory filter functions under short and long intervals. 
Uniform weighting performs poorly in pure saccade behavior scenarios ($G_4$). During saccades, however, later gaze points are generally more informative than earlier ones, and uniform weighting therefore overemphasizes irrelevant early gaze samples.
Linear weighting improves performance by prioritizing later gaze points, but under long intervals with saccade behavior ($G_4$) it still assigns excessive weight to early samples, leading to degraded accuracy. 
 OPTICS clustering requires a sufficient number of gaze samples and therefore fails in short interval scenarios with saccade behavior ($G_2$-$G_4$), while the last-gaze strategy degrades in saccade cases ($G_3$ and $G_4$) where the later gaze point does not reliably reflect user intention (last gaze noise, as shown in Fig. \ref{weight} (d)). 
Linear weighting and our method without OPTICS clustering perform poorly in the long interval $G_3$ scenario (fixation followed by gaze drift). In this special case, early gaze samples are more informative than later ones, and the lack of explicit fixation–saccade differentiation leads to incorrect emphasis on less relevant gaze points.  
In contrast, the proposed method explicitly models fixation and saccade behaviors and adaptively balances early and late gaze contributions, achieving robust performance across different intervals and gaze patterns.}

\subsubsection{\textcolor{black}{Ablation study of command processing}}

\begin{table}[t]

\caption{\textcolor{black}{Ablation study of command processing.}}
\vspace{-10pt}
\label{ablation command processing}
\begin{center}
\setlength{\tabcolsep}{5pt}
\begin{tabular}{ccc}
\hline
\hline
\multicolumn{1}{c}{\multirow{2}{*}{Approach}} &  \multicolumn{2}{c}{\shortstack{\\Success rate $\uparrow$ }}  \\ \cmidrule(r){2-3} 
 &     Rule based  &  Fuzzy \\ \hline
 \multicolumn{3}{c}{\multirow{1}{*}{\textit{\textbf{Human View Command Processing}}}}  \\
Keyword trigger & 0.94 & 0  \\
Our w. fixed time window ($N=3$)  & 0.81  & 0.79 \\  
Our w. fixed time window ($N=6$)  & 0.76  & 0.71  \\ \hline
  \multicolumn{3}{c}{\multirow{1}{*}{\textbf{\textit{Planning Policy Generation}}}} \\
Behavior trees planner& 0.94 & 0 \\ \hline
Our & \textbf{0.98}  &  \textbf{0.91} \\
\hline 
\hline
\end{tabular}   
\end{center}
\footnotesize{ \textcolor{black}{\textbf{Remark: }
Fuzzy command: The manipulation-related verb and keywords are not explicitly included, e.g., I want to eat the fruit with the plate. Rule-based command: Command explicitly contains keywords and manipulation-related verbs (not natural sentences) \cite{c17}, e.g., pick this, put there.}

}

\end{table}

\textcolor{black}{
Tab.~\ref{ablation command processing} presents an ablation study of the two LLM agents for command processing. 
  For human-view command processing, a keyword-trigger baseline replaces the LLM agent. This baseline relies on predefined keywords and heuristic rules~\cite{c17}.
While effective for rule-based commands with explicit manipulation verbs, it fails under fuzzy commands where no clear keywords are available. 
We further evaluate fixed time-window variants, where gaze samples are aggregated with speech command using a predefined window length.
However, short time windows suffer from insufficient gaze samples, whereas long time windows introduce excessive noise, leading to degraded performance under both command types.
}

\textcolor{black}{
For planning policy generation, replacing the LLM agent with a behavior-tree planner yields strong performance for rule-based commands but fails under fuzzy commands due to its dependence on explicit command structures. }

\textcolor{black}{In contrast, our approach maintains high success rates across both rule-based and fuzzy commands by avoiding reliance on predefined keywords or rigid command templates, enabling more robust natural language interaction.
}

\subsubsection{\textcolor{black}{Ablation Study of Multi-view Object Alignments}}

\begin{table}[t]

\caption{\textcolor{black}{Ablation study of multi-view object alignments}}
\vspace{-10pt}
\label{ablation multiview}
\begin{center}
\setlength{\tabcolsep}{5pt}
\begin{tabular}{ccccc}
\hline
\hline
\multicolumn{1}{c}{\multirow{2}{*}{Method}} &  \multicolumn{4}{c}{\shortstack{\\Correct Alignment Probability $\uparrow$ }}  \\ \cmidrule(r){2-5} 
 &   $45^{\circ}$ & $90^{\circ}$ & $135^{\circ}$& $180^{\circ}$\\ \hline
ARCalibration \cite{9889538} & 0.88  & 0.74 & 0.91 & 0.82   \\ \hline
Our & \textbf{0.99}  & \textbf{ 0.93} & \textbf{ 0.92} & \textbf{ 0.96} \\ 
\hline 
\hline
\end{tabular}   
\end{center}
\footnotesize{\textcolor{black}{\textbf{Remark:} Correct Alignment Probability = correct selected referred object / number of trials.  The angle denotes the relative orientation between the two cameras projected onto the tabletop plane.}

}

\end{table}

\textcolor{black}{Tab.~\ref{ablation multiview} reports an ablation study of multi-view object alignment under varying relative viewpoints between the RGB camera on ARIA glasses and the robot camera.}

\textcolor{black}{We compare our feature matching based alignment method with ARCalibration~\cite{9889538}, which estimates the headset pose using multiple QR codes.  However, QR codes cannot be detected at every frame, resulting in reduced valid gaze points and accumulated gaze drift.}

\textcolor{black}{In contrast, our method leverages object-level feature matching between human and robot views and does not rely on continuous detection of calibration markers. As a result, it maintains consistently high alignment accuracy across a wide range of viewing angles. These results demonstrate the robustness of the proposed multi-view alignment strategy under a large viewpoint difference.}

\subsubsection{\textcolor{black}{Ablation Study of Modalities}}

\begin{table}[t]

\caption{\textcolor{black}{Ablation study of modalities}}
\vspace{-10pt}
\label{abaltion modalities}
\begin{center}
    \centering
    \begin{tabular}{ccc}
     \hline
      \hline
     Modality  &  Success Rate $\uparrow$  & Interaction time $\downarrow$ (s)
     \\ \hline
     Speech only
    & 0.90& 7.3  \\ \hline
    Gaze only 
    & 0.65& 23.7 \\ \hline
   Our
   & \textbf{0.94} & \textbf{2.2}\\ 
    \hline
    \hline
    \end{tabular}
    \end{center}
    \footnotesize{\textcolor{black}{\textbf{Remark:} The experiment was conducted in the $S_3$ (Multi-Step Action).} }
    \vspace{-10pt}
\end{table}

\textcolor{black}{Table~\ref{abaltion modalities} presents an ablation study evaluating the contribution of interaction modalities.
Using speech alone leads to increased interaction time, as users must verbally disambiguate similar objects. 
Gaze-only interaction performs worst in both success rate and interaction time. Without language input, users are required to maintain long fixation on target objects, and action selection must rely on additional GUI, which increases cognitive load and interaction duration and results in frequent failures.}

\textcolor{black}{In contrast, the proposed multimodal approach combines the complementary strengths of speech and gaze. Speech provides high-level task semantics, while gaze offers rapid and intuitive object disambiguation. Our proposed approach enables both higher success rates and significantly shorter interaction times, demonstrating the effectiveness of human-robot collaboration.}

  \section{Conclusion}

We proposed FAM-HRI, a multimodal HRI framework that fuses real-time gaze and speech inputs from ARIA glasses with robot vision using LLM-based agents. Our system integrates human intention fusion, which identifies the target object and the relevant fixation period, with a multi-view alignment strategy that reconciles observations from the human and robot perspectives. The planning policy generation module then produces robust, parameterized action primitives for precise and efficient task execution in dynamic and cluttered environments. Experimental evaluations and user studies confirm the improved interaction performance offered by our approach. 

\section{Limitations and Future Works} 
FAM-HRI has several limitations. First, latency from LLM and VLM inference may
affect real-time responsiveness and make deployment on edge
devices challenging. Future work will explore accuracy-speed trade-offs, such as
fine-tuning parameter-efficient models with Low-Rank Adaptation (LoRA)
\cite{hu2022lora} and implementing retrieval-augmented generation (RAG)
\cite{yuan2026rpms} with smaller backbones to reduce latency.

Second, loose-fitting ARIA glasses may cause gaze-estimation misalignment.
Cluttered or outdoor environments can further introduce uncertainty through
lighting changes, occlusions, and background motion. To improve robustness,
future work will focus on continuous time formation \cite{nguyen2026third} or a VLM that explicitly incorporates diverse
environmental conditions during training \cite{Yang_2025_CVPR}, enabling better generalization across real-world HRI scenarios.

Finally, our current framework does not explicitly handle rare cases where speech and gaze refer to different objects, e.g., when the user looks at a cup on the left while verbally referring to a similar cup on the right. The current method resolves such ambiguity by selecting the object with the lowest weighted distance to the gaze trajectory. Although such conflicts are uncommon in practice, future work will address them by integrating a dialogue module \cite{c21}, allowing the robot to clarify user intent, for example by hovering above a candidate object before execution. This would improve disambiguation in more complex multimodal contexts.

\bibliographystyle{IEEEtran}
\bibliography{mybib}

\appendices

\section{Evaluation of Tabletop Task }
\label{Evaluation of Tabletop Task}
\begin{table*}[b]    

\caption{\textcolor{black}{Detailed simulation tabletop manipulation success rate (\%) and interaction time (s) across different task scenarios}}
\begin{center}
    \centering
    \begin{tabular}{lccccc}
     \hline
      \hline
         & \multicolumn{1}{c}{\multirow{2}{*}{\shortstack{\\Success Rate (\%)}}} & \multicolumn{1}{c}{\multirow{2}{*}{\shortstack{\\Interaction time (s)}}} & \multicolumn{1}{c}{\multirow{2}{*}{\shortstack{\\Number of  \\parameters}}}   & \multicolumn{1}{c}{\multirow{2}{*}{\shortstack{\\Number of \\ actions}}} & \multicolumn{1}{c}{\multirow{2}{*}{\shortstack{\\Complexity }}}  \\ 
         \\ 
         \hline
        \textit{ \textbf{With referred object}} & & \\
    pick up the $<$object$>$    &100 & 1.3 & 1 & 1 & 2 \\
    grab the pieces ($S_1$)     &96 & 1.8 & 1 & 1 &2\\ 
    put the $<$object$>$ on the $<$plate$>$ &100 & 2.2 &2 & 3 & 5\\

    \multicolumn{1}{l}{\multirow{2}{*}{\shortstack[l]{\\put the $<$object$>$ on the $<$plate$>$ then \\ pour something from the $<$cup$>$ on it}}}    & \multicolumn{1}{c}{\multirow{2}{*}{\shortstack{\\94}}}&\multicolumn{1}{c}{\multirow{2}{*}{\shortstack{\\5.9}}}  &\multicolumn{1}{c}{\multirow{2}{*}{\shortstack{\\4}}}&\multicolumn{1}{c}{\multirow{2}{*}{\shortstack{\\6}}}&\multicolumn{1}{c}{\multirow{2}{*}{\shortstack{\\10}}} \\ \\ \\
    put this $<$object$>$ $<$there$>$ (Position on table) &  89 & 2.5&2&3&5 \\ 
    put the $<$object1$>$ and $<$object2$>$ on the $<$plate$>$ & 92 &3.8 &3&6&9\\ 
     \multicolumn{1}{l}{\multirow{2}{*}{\shortstack[l]{\\put the $<$object1$>$ on the $<$plate1$>$ then \\ put the $<$object2$>$ on the $<$plate2$>$}}}     &\multicolumn{1}{c}{\multirow{2}{*}{\shortstack{\\90}}} & \multicolumn{1}{c}{\multirow{2}{*}{\shortstack{\\6.1}}}&\multicolumn{1}{c}{\multirow{2}{*}{\shortstack{\\4}}}&\multicolumn{1}{c}{\multirow{2}{*}{\shortstack{\\6}}}&\multicolumn{1}{c}{\multirow{2}{*}{\shortstack{\\10}}} \\ \\ \\
      \multicolumn{1}{l}{\multirow{2}{*}{\shortstack[l]{\\grab the $<$object$>$ and lift up for $<$distance$>$ then \\ turn it for $<$angle$>$ degrees}}}      &  \multicolumn{1}{c}{\multirow{2}{*}{\shortstack{\\96}}} & \multicolumn{1}{c}{\multirow{2}{*}{\shortstack{\\8.3}}}&\multicolumn{1}{c}{\multirow{2}{*}{\shortstack{\\3}}}&\multicolumn{1}{c}{\multirow{2}{*}{\shortstack{\\3}}}&\multicolumn{1}{c}{\multirow{2}{*}{\shortstack{\\6}}}                   \\ \\   \\ \hline
    \textit{\textbf{Without referred object}} &  & \\
    pick up this    & 100 & 1.1 & 1 & 1 &2 \\
    grab this ($S_1$)    & 87 & 0.8 &1 &1&2 \\ 
    put this on that & 98 & 1.7&2 &3&5\\ 
    put this on that then pour something from this on it   & 96  & 4.2 &4&6&10\\ 
    put this $<$there$>$ (Position on table) & 93 &1.9&2&3&5 \\ 
    put this and this on that & 96 &3.1&3&6&9 \\ 
    put this on this then put this on that  &92 &4.2&4&6&10 \\ 
    \multicolumn{1}{l}{\multirow{2}{*}{\shortstack[l]{\\grab this and lift it up for $<$distance$>$ then \\ turn it for $<$angle$>$ degrees}}}  &  \multicolumn{1}{c}{\multirow{2}{*}{\shortstack{\\95}}} &\multicolumn{1}{c}{\multirow{2}{*}{\shortstack{\\7.9}}}   &\multicolumn{1}{c}{\multirow{2}{*}{\shortstack{\\3}}}&\multicolumn{1}{c}{\multirow{2}{*}{\shortstack{\\3}}}&\multicolumn{1}{c}{\multirow{2}{*}{\shortstack{\\6}}}           \\ \\ \\    \hline
    \textbf{\textit{Fuzzy Command}} \\
    Show me the $<$object$>$ &96 & 1.6 & 1 & 1 & 2 \\
    Show me $<$that$>$ & 94 & 1.4 & 1 & 1 & 2 \\
    I want to eat $<$this$>$ with the $<$plate$>$ & 93 & & & & \\
     \multicolumn{1}{l}{\multirow{2}{*}{\shortstack[l]{\\ I'm hungry, I want some $<$fruit$>$, I think this \\ is better, I want to eat with the $<$plate$>$}}} & \multicolumn{1}{c}{\multirow{2}{*}{\shortstack{\\91}}} & \multicolumn{1}{c}{\multirow{2}{*}{\shortstack{\\8.3}}} & \multicolumn{1}{c}{\multirow{2}{*}{\shortstack{\\2}}} & \multicolumn{1}{c}{\multirow{2}{*}{\shortstack{\\3}}} & \multicolumn{1}{c}{\multirow{2}{*}{\shortstack{\\5}}} \\ \\ \\ 
   \multicolumn{1}{l}{\multirow{2}{*}{\shortstack[l]{\\ I want to eat this using the $<$plate$>$, \\ some cream from the $<$cup$>$ on it would be better}}}    & \multicolumn{1}{c}{\multirow{2}{*}{\shortstack{\\83}}} &  \multicolumn{1}{c}{\multirow{2}{*}{\shortstack{\\9.2}}} & \multicolumn{1}{c}{\multirow{2}{*}{\shortstack{\\4}}} & \multicolumn{1}{c}{\multirow{2}{*}{\shortstack{\\6}}} & \multicolumn{1}{c}{\multirow{2}{*}{\shortstack{\\10}}}\\ \\ \\
    \hline
    \hline
    \end{tabular}
    \end{center}
    \footnotesize{ \textbf{Remark:} The listed commands are indicative examples. Our system supports a wide range of natural language expressions with similar semantic intention, allowing flexible and intuitive user interaction beyond fixed templates.}
    \label{tab:task}
\end{table*}

\textcolor{black}{We define task complexity as the sum of (1) the number of predefined actions in the action primitives and (2) the number of parameters, where parameters encompass both the number of referred objects and the associated spatial specifications such as positions and angles.}

\textcolor{black}{Tab. \ref{tab:task} presents the success rates and average interaction times for various tabletop manipulation tasks. A key observation is that tasks executed without a referred object generally exhibit faster interaction times due to shorter command lengths. However, in Scenario $S_1$, the success rate for commands without explicit object references is lower. This is primarily because small, detailed parts of the chess pawns were sometimes misdetected by VLM. Consequently, during multi-view alignment, insufficient keypoints in these small detected parts were found, reducing the accuracy of object selection. Additionally, as the length and complexity of the user's command increase, the reasoning process required by the two LLM-Agents becomes more involved, which in turn leads to a decrease in success rate due to higher inference uncertainty. Our system maintains robustness when encountering fuzzy commands lacking specific verbs.}

\section{Set up for Noise Environment}
\label{Noise Environment}
\begin{figure}[t]
      \centering
      \includegraphics[width=0.48\textwidth]{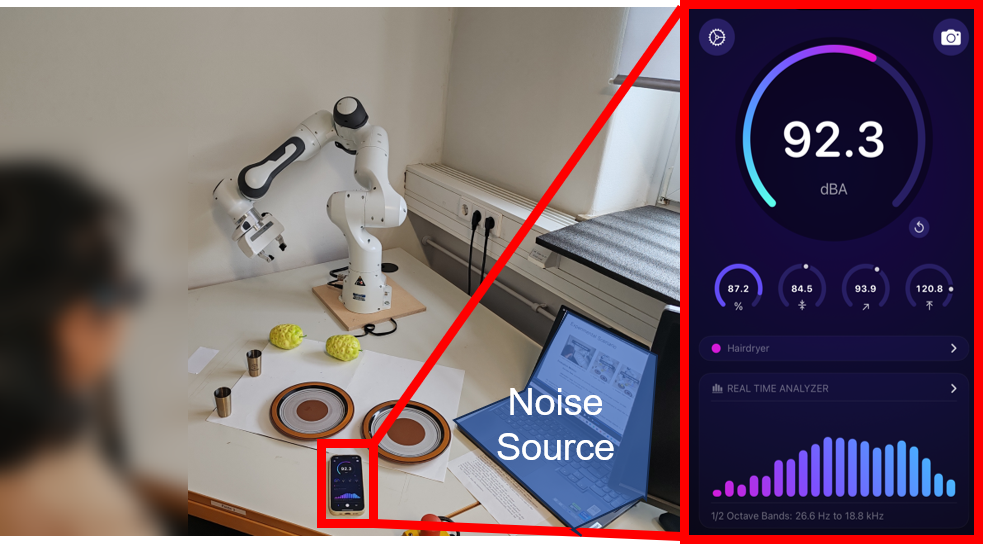}
      \caption{System robustness evaluation under varying background noise. The system consistently achieved 100\% success rate up to 70    dB.}
      \label{noiseenv}
\end{figure} %
To evaluate the robustness of FAM-HRI in real-world conditions, we conducted experiments under varying ambient noise levels. 
As shown in Fig. \ref{noiseenv}, the background noise was introduced using a laptop placed near the user, playing a pre-recorded dialog. The content of the recording simulated a real indoor conversation scenario, ensuring that the noise characteristics are similar to those encountered in household or public settings. A sound level meter was positioned next to the user to measure the real-time noise intensity. 
\section{\textcolor{black}{Sensitivity Analysis of the Weight Factor}}
\label{Sensitivity}
  \begin{figure}[t]
      \centering
      \includegraphics[width=0.4\textwidth]{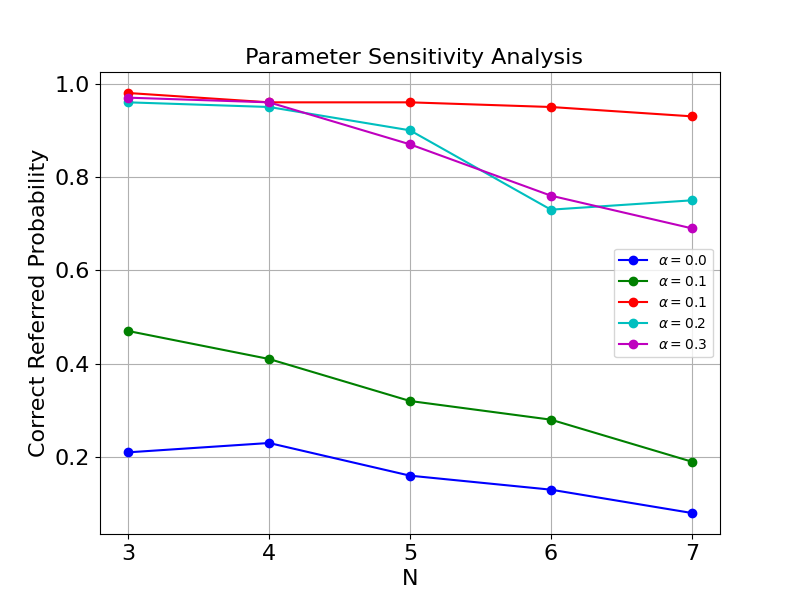}
     
      \caption{\textcolor{black}{Sensitivity Analysis of the Weight Factor}}
      \label{sens}
      
  \end{figure}

\textcolor{black}{Fig.~\ref{sens} shows the sensitivity of the weighting factor $\alpha$ under the $G   _4$ scenario (as shown in Fig. \ref{G1-3}). 
When $\alpha$ is small, the weighting function approaches uniform weighting, causing irrelevant early gaze samples to dominate and resulting in poor performance. 
Increasing $\alpha$ improves intention recovery by emphasizing later gaze points. 
However, for long gaze trajectory, overly large $\alpha$ overweights the later gaze point, which may not correspond to user intention, leading to performance degradation. 
These results indicate that a moderate $\alpha$ provides the best balance between robustness and sensitivity.}

\section{\textcolor{black}{ Prompt Engineering for FAM-HRI}}
\label{Prompt}

\subsection{\textcolor{black}{LLM-Agent for Human View Command Processing}}
\label{PromptA}
\textcolor{black}{Given a single user command, we infer the user’s overall intent and output the set of physical targets a robot must manipulate to satisfy that intent (manipulable objects or explicit spatial positions). For each target, we also output one word in the command that is most likely spoken while the user is gazing at that target. The prompt rules are as follows:}

\begin{enumerate}
 
    \item Goal Inference: Infer the user’s intention from the entire expression.
Identify which target object must be involved for the intention.
\item Target Enumeration: Output only physical, robot-relevant targets (manipulable objects, tools, source containers, or explicit positions).
\item Pronoun Resolution: If the command contains an explicit manipulable object mention that can serve as an antecedent, resolve the pronoun to that object. If no explicit antecedent exists in the command, set ``$object\_name$"=``$stuff$".
\item  Gaze–Word Alignment: For each target you output, provide one word in the command at which the user is most likely looking at that target.
\item Location \& Deictic-Position Rule: If deictic location words indicates a placement destination, label the target as ``$Character$"=``$position$". If an anchoring phrase is present (e.g., ``there on the table”), set "$object_name$"="$table$" (the anchoring object must be detected) while the gaze-anchor word remains there.
\item  Target Merging: If the same target is referred to multiple time, you must merge them into one entry 
\item Output Format: Return only one JSON object.

\end{enumerate}

\subsection{\textcolor{black}{LLM-Agent for Planning Policy Generation}}
\label{PromptB}
\textcolor{black}{Central to planning policy generation of our FAM-HRI is an LLM generating JSON object. We did not have the LLM generate Python code directly but instead had it generate JSON objects, as this approach is safer and facilitates easier executable review. We provide the following
environment APIs that LLMs can choose to invoke:}

\begin{itemize}

    \item{} Pick: Grip an object at the specified ``${position}$". This action should be followed by specifying the position within a 2D list. e.g., ``pick": [[aaaa, bbbb]].
    \item{} Put: Release an object at the specified ``${position}$". This action requires a subsequent 2D list to indicate the position of the object, e.g. ``put": [[cccc, dddd]]
    \item {} Pour: Pour something on the specified ``$2d_{position}$. This action requires a subsequent 2D list to indicate the position of the object, e.g. ``pour": [[ccc, dddd]]. Do not open the gripper after pour action.

\item {} MoveTo: Navigate to a target ``${position}$". This action requires a subsequent 2D list to indicate the position of the object, e.g. ``MoveTo": [[cccc, dddd]].
\item {} Swap: Swap the positions of two objects ``${positionA}$" and ``${positionB}$". This action requires two 2D lists to indicate the position of the object, e.g., ``Swap": [[cccc, dddd]], [[aaaa, bbbb]].

\item {} MoveXYZ: Move the end-effector
along the x, y, or z axes. Using a two-dimensional array ``[x, y, z]" in meters, where x is positive for up, negative for down, y is positive for right and negative for left, z is positive for forward and negative for backward. Use this action only when the user specifies that the robot arm needs to be moved with a specific distance.
\item {} OpenGripper: Open the robot's gripper. 

\item {} CloseGripper: Secure an object by closing the gripper. 
\item {} Rotate: Change the gripper's orientation for precise alignment. This action should be followed by specifying the angle within a 1D list, e.g. ``rotate": [[aaa]]

\end{itemize}

\section{\textcolor{black}{Experimental Setup for User Study}}
\label{userstudy}
\begin{figure}[t]
      \centering
      \includegraphics[width=0.4\textwidth]{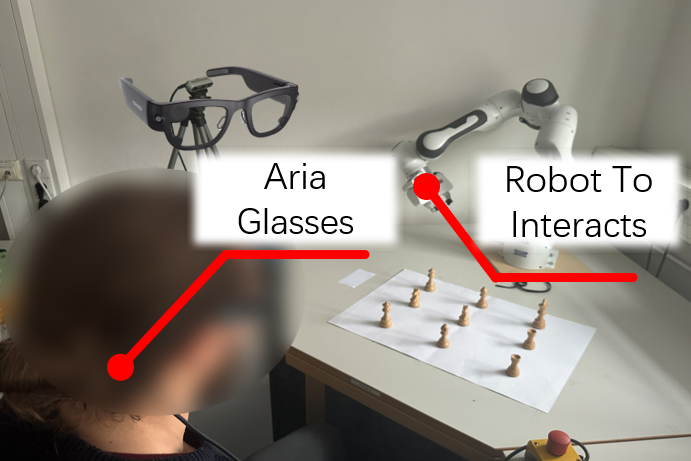}
      \caption{\textcolor{black}{Experimental Setup for user study. A participant equipped with Aria glasses issues commands to the robot manipulator.}}
      \label{figuredisable}
 \end{figure}

\textcolor{black}{The experimental setup for user study is shown in Fig. \ref{figuredisable}. All participants are  fluent in English. Among the German participants, none have physical impairments; the group includes 4 women and 14 men, with one individual over the age of 60 and the rest under 50. Among the Singaporean participants, \textcolor{black}{20} have mild upper-limb impairments; the group consists of \textcolor{black}{8} women and \textcolor{black}{12} men, with two individuals over the age of 60 and the rest under 50. Each individual received only brief instructions and was asked to naturally move their eyes among the multiple referred objects during the experiment. }

\section{Failure Modes}

\begin{figure}[t]
      \centering
      \includegraphics[width=0.45\textwidth]{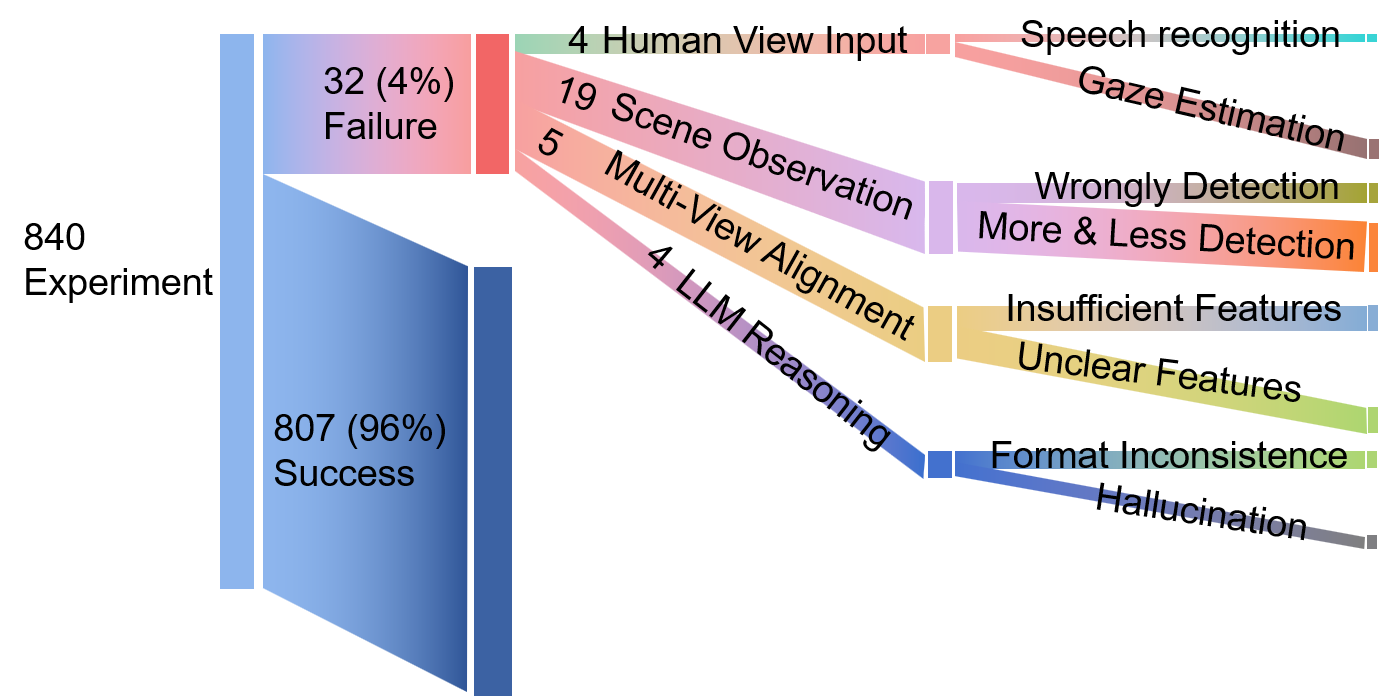}
     
      \caption{Failure Modes.}
      \label{failure}
\end{figure} %

As shown in Fig. \ref{failure}, the failure cases of our system can be categorized into four key aspects: human view input, scene observation, multi-view alignment, and LLM reasoning.
\begin{enumerate}
    \item \textbf{Human View Input:} For human view input, there are two sources of failure, speech recognition errors, and inaccurate gaze estimation. The speech recognition error in our system is primarily due to misinterpretation of similar-sounding words. Gaze estimation errors are largely influenced by improper wear of the glasses. Specifically, when the nose bridge of the ARIA glasses is not correctly placed on the user's nose, the accuracy of gaze estimation decreases significantly. However, to mitigate this, users were instructed to adjust their glasses properly before proceeding with the experiment, ensuring that such errors do not propagate into the system. 
    \item \textbf{Scene Observation:} For scene observation, failures come mainly from more or less detected objects. In Scenario $S_1$, where users need to select a specific chess piece without explicitly stating its category, smaller parts of the chess pieces were sometimes incorrectly detected as separate objects. Additionally, speech recognition errors could result in misclassification of the target object, causing the system to detect an incorrect category and subsequently select the wrong object.
    \item \textbf{Multi-View Alignment:} For multi-view alignment, the main reason for failure is insufficient or unclear feature correspondence between human and robot views. Feature matching using superglue becomes unreliable in the presence of weak object textures, repetitive patterns, or partial occlusions, leading to incorrect correspondences.
    \item \textbf{LLM reasoning:} For LLM Reasoning, the primary failure modes come from formatting inconsistencies and hallucinations in the generated output. Since FAM-HRI requires the LLM’s response to strictly follow a predefined prompt format, any deviation renders the output unusable by subsequent system modules. In rare cases, the LLM misinterprets the required action parameters, leading to incorrect parameter values or incorrect ordering of action arguments. Additionally, logic errors occur in policy generation due to LLM hallucinations. For example, in placement tasks, the generated policy may not open the gripper after reaching the target position, leading to task failure. However, such errors were observed infrequently in our experiments, as structured prompt design and output constraints significantly reduced the occurrence of these issues.
\end{enumerate}

\section{\textcolor{black}{Demonstration of Deterministic Conflict Resolution Strategy}}
\label{demon}

\begin{figure*}[t]
      \centering
      \includegraphics[width=0.99\textwidth]{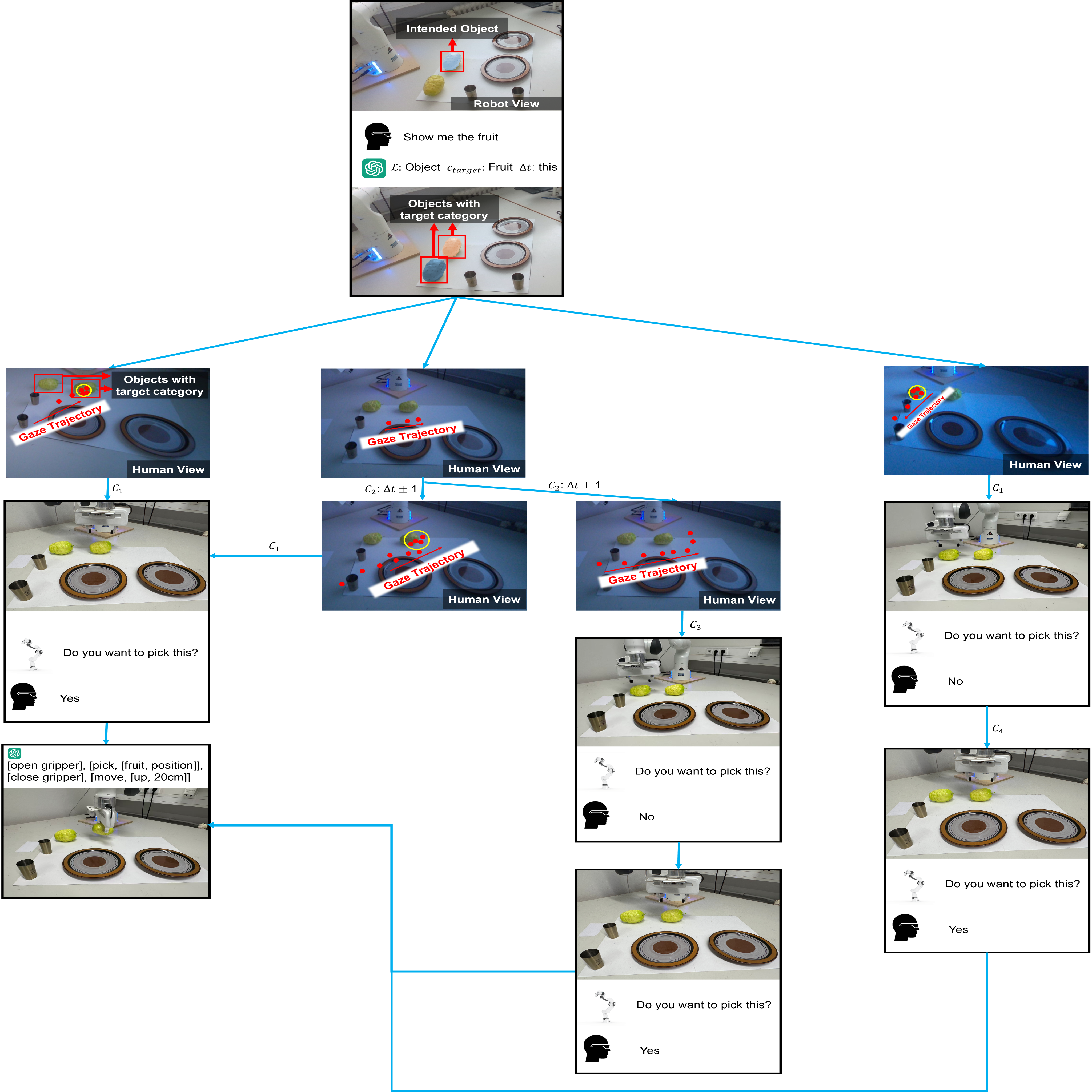}
      \vspace{-5pt}
      \caption{ \textcolor{black}{Demonstration of Deterministic Conflict Resolution Strategy. The user wants the robot to pick up the fruit, which is on the top from the robot's view and which is on the right from the human's view.}}
      \label{figdemon}
 
\end{figure*} %

\section{\textcolor{black}{Data Collection and Logging Procedure}}

\textcolor{black}{All user data is handled confidentially; in particular, images containing identifiable facial information and pupil images are not stored. During data collection, we log each modality and intermediate output in a structured format: speech is saved as a CSV file containing the start and end timestamps of each word; gaze points are stored as a Python list of 2D points; images are saved as PNG files; object detection results are stored as JSON files; and the LLM outputs are saved as TXT files. This makes the data pipeline transparent and reproducible. }

\section{\textcolor{black}{Analysis of Variance}}
\label{ANOVA}

\textcolor{black}{We mapped five-point Likert responses from ``strongly disagree" to ``strongly agree" to numerical scores from $-2$ to$+2$ (higher is better). For each dimension, we report $mean\pm std$  and compared Normal vs. Disabled participants using a one-way ANOVA.}

\textcolor{black}{Tab. \ref{tabanova} shows that there are no statistically significant differences between Normal and Disabled participants on the Accurate, Flexible, Modern, or Efficient dimensions (all p $>$
0.05), indicating broadly consistent evaluations across groups; however, a significant group effect is observed for Simple (p = 0.003), with Disabled participants rating the system as simpler than Normal participants (1.72 vs. 0.74).}

\begin{table}[t]

\caption{\textcolor{black}{Results of Analysis of Variance.}}
\label{tabanova}
\begin{center}
\setlength{\tabcolsep}{5pt}
\begin{tabular}{cccc}
\hline
\hline
\multicolumn{1}{c}{\multirow{1}{*}{Dimensions}} &  \multicolumn{1}{c}{\shortstack{\\Normal Participants }} &  \multicolumn{1}{c}{\shortstack{\\Disabled participants }} & $P$   \\ \hline
Accurate& $0.72\pm 1.03$ & $0.85\pm 1.26$ & $0.734$ \\ 
Flexible & $0.36\pm 1.13$ & $0.65\pm 1.08$ & $0.409$ \\ 
Modern& $0.68\pm1.42$ & $1.15\pm 1.22$ & $0.260$\\ 
Simple & $0.74\pm 1.35$ & $1.72\pm 0.63$ & $0.003$\\ 
Efficient & $1.54\pm 0.67$ & $1.65\pm 0.48$ & $0.565$ \\ 
\hline 
\hline
\end{tabular}  
\end{center}
\footnotesize{\textcolor{black}{\textbf{Remark:} Participants evaluated the proposed system along five dimensions: \emph{Accurate}, \emph{Flexible}, \emph{Modern}, \emph{Simple}, and \emph{Efficient}.
}

}

\end{table}

\section{Parameter Tuning and Reproducibility}
\begin{table}[t]
\footnotesize
    \centering
    \renewcommand{\arraystretch}{1.5}
    \caption{List of Parameters}
    \begin{tabular}{cc}
        \hline
         \hline
\textbf{Grounding Dino SAM2} & \\
Box Threshold &0.3  \\
Text Threshold & 0.3 \\
Grounding-Model & Tiny \\
SAM2 Model & Large \\ \hline
\textbf{SuperGlue} & \\
Max Keypoints & 10000 \\
Keypoints Threshold & 1e-5 \\
Match Threshold & 1e-5 \\
Resize & No \\
Weights & indoor \\
        \hline
         \hline
    \end{tabular}
    
    \label{tab:re}
\end{table}
Our approach involves several parameters that are manually set or empirically tuned based on experimental observations. The Weight Factor in intention alignment was determined through extensive testing to balance the influence of recent gaze points while preventing a single point from dominating the selection process. Although these parameters have been carefully adjusted to optimize system performance, we acknowledge that they are not mathematically proven to be optimal.

To ensure reproducibility, we provide explicit parameter settings for key components of our system in Tab. \ref{tab:re}, including Grounding Dino for Scene Observation and SuperGlue for Multi-View Alignment.

\section{List of symbols and their meanings}
\label{symbols}

To provide a comprehensive reference for the mathematical notation used throughout our paper, we include Tab. \ref{tab:symbols}. This table details all symbols, spaces, sets, and corresponding meanings, covering essential aspects of frame transformations, gaze reconstruction, scene observations, intention fusion, control system variables, and policy generation.

The notation system is structured to facilitate clarity in understanding multi-view geometric transformations, gaze-language fusion, and LLM-based policy generation. Specifically, it includes:
\begin{enumerate}
    \item Frame representations for robot base, cameras, and gaze.
\item Scene Observations for both human and robot views, ensuring a unified representation for multi-view alignment.
\item Control system parameters, including action primitives, parameters, and planning policies.
\end{enumerate}

This structured notation serves as a foundation for understanding FAM-HRI’s multimodal interaction pipeline, enabling reproducibility and further development within the research community.
\begin{table*}[t]
\footnotesize
    \centering
    \renewcommand{\arraystretch}{1.5}
    \caption{List of symbols and their meanings}
    \begin{tabular}{ccc}
        \hline
         \hline
        \textbf{Symbol} & Space & \textbf{Meaning} \\
        \hline

        $\prescript{r}{}{()}$ &  -   & Frame: Robot Base   \\
        \hline

        $\prescript{c}{}{()}$ &  -   & Frame: Robot Camera   \\
        \hline

        $\prescript{gc}{}{()}$ &  -   & Frame: Glasses Camera   \\
        \hline
        $\prescript{gp}{}{()}$ &  -   & Frame: Glasses Pupil   \\
        \hline

        $\mathcal{K}$ & $\mathbb{R}^{3 \times 3}$ & Basic Geometry: Camera Intrinsic Matrix  \\
        \hline

        ${p}$ & $\mathbb{R}^2$ &  Basic Geometry:  2D Position of a 3D coordinate  \\
        \hline
       
        $\mathbf{P}$ & $\mathbb{R}^3$ &  Basic Geometry:   3D coordinate  \\
        \hline

        ${}^{gc}T_{gp}$ & $\mathbb{SE}(3)$ & Transformation from Glasses Camera Frame to Glasses Pupil Frame  \\
        \hline

       ${}^{s}T^{head}_{t_i}$ & $\mathbb{SE}(3)$ & Transformation Representing Head Pose at Time $t_i$  \\
        \hline

        $\beta$ &  $\mathbb{R}^{4}$ & Environmental Bounding Box Representation $\left( [x_{\min}, y_{\min}, x_{\max}, y_{\max}] \right)$  \\
        \hline
        
        $M$ &  $\mathbb{R}^{H \times W}$ & Environmental Segmentation Mask (binary or probability map)  \\
        \hline

        $\mathcal{S}$ & $\Sigma^*$  & Human Input Speech Sequence in Text \\
        \hline
        
        $\mathcal{G}$ & $\mathbb{R}^{H \times W}$ & Human input of Eye tracking camera   \\
        \hline

        $\mathcal{U}$ & $\mathbb{R}^{H \times W*(t_n-t_1)}$ & Human view: Image from User-Centric Camera on glasses   \\
        \hline

        $\mathcal{Z}_{H}$ & $\Sigma^* \times \mathbb{R}^{H_1 \times W_1 \times (t_n - t_1)} \times \mathbb{R}^{H_2 \times W_2 \times (t_n - t_1)}$  & Human Input State Vector / Descriptor  \\
        \hline

        ${}^{gc_{t_{i}}} \mathbf{P}_{{t_{i}}}$ & $\mathbb{R}^3$ & Human view: 3D Gaze Vector at Time $t_i$ in Glasses Carmea Frame at Time $t_i$ \\
        \hline
  ${}^{gc_{t_{i}}} \mathbf{P}_{{t_{i+n}}}$ & $\mathbb{R}^3$ & Human view: 3D Gaze Vector at Time $t_{i+n}$ in Glasses Carmea Frame at Time $t_i$ \\
        \hline
 $\mathcal{Z}_{g}$ &   &  Human view: Scene Observations Set   $\left \{ c_{target}, \ \{{p}_{g_i}, \ \beta_{g_i}, \ M_{g_i} \}_{i=1}^{N} \right \}$  \\
        \hline

 ${p}^{gaze}$ & $\mathbb{R}^2$ & Human View: 2D Projected Gaze Point on the Image Plane of $\mathcal{U}$  \\
        \hline
     ${}^{gc_{t_{i}}} {p}^{gaze}_{{t_{i+n}}}$ & $\mathbb{R}^2$ & Human View: Projected Gaze Point at Time $t_{i+n}$ on Image at Time $t_{i}$  \\
        \hline

    $\overline{\mathcal{Z}_{g}}$ & $( \overline{{p}_{g}}, \  \overline{\beta_{g}}, \  \overline{M_{g}} )$ & Human view: Referred Object   \\
        \hline

        $\mathcal{C}$ & $\mathbb{R}^{H \times W}$ & Robot view: Image from Camera of Robot   \\
        \hline
        
        $\mathcal{Z}_r$ &  & Robot view: Scene Observations Set   $\left \{ c_{target}, \ \{{p}_{r_i}, \ \beta_{r_i}, \ M_{r_i} \}_{i=1}^{N} \right \}$ \\
        \hline
  $\overline{\mathcal{Z}_{r}}$ & $( \overline{{p}_{r}}, \  \overline{\beta_{r}}, \  \overline{M_{r}} )$ & Robot view: Referred Object   \\
        \hline

     $\mathbf{O}_1$ & $\left \{  \mathcal{L},\ c_{target},\ \Delta t   \right \} $ & Control system: LLM Output for Gaze Time Period Prediction  \\
        \hline
  $\mathcal{L}$ & $\Sigma^*$ & Control System: Target Property Descriptor (e.g., Object, Position)  \\
        \hline

        $c_{target}$  & $\Sigma^*$ & Control System: Target Object Category from User Command  \\
        \hline

        $\Delta t$ & $\mathbb{R}$ & Control System: Time Period of User's Gaze Fixation on Target Object  \\
        \hline

       $X$ & $\left \{ \overline{\mathcal{Z}_{r}}, \ \mathbf{O}_1, \ \mathcal{S} \right \} $    & Control System: State Space   \\
        \hline

        $\theta$ & $ \mathbb{R}^n$ & Control System: Action Parameter Space (Rotation, Translation, position, etc.)   \\
        \hline

        $a$ &  $\Sigma^*$ & Control System: Action Space Primitives   \\
        \hline

        $\mathcal{A}$ & $\left \{ a(\theta)  \right \} $   & Control System: Set of Action Primitives   \\
        \hline

        $\alpha$ & $\mathbb{R}$ & Weighted Decay Factor for Temporal Gaze Analysis  \\
        \hline

        $\pi$ & $ \pi : X \mapsto \ \mathcal{A}\times \Theta$ & Control System: Parameterized  Planning Policy for Executing Actions   \\
        \hline
         \hline
    \end{tabular}
    
    \label{tab:symbols}
\end{table*}

\end{document}